\documentclass{andp2012}

\usepackage{epsfig}
\usepackage{graphicx}
\usepackage{subfigure}
\usepackage{color}
\usepackage{mathrsfs}
\usepackage{amsfonts}
\usepackage[english]{babel}

\makeatletter
\renewcommand{\@thesubfigure}{\hskip\subfiglabelskip}
\makeatother

\category{Original Paper}
\keywords{Quantum phase transitions, spin boson model, variational method.}
\title{Variational study of two-impurity spin-boson model with a common Ohmic bath: Ground-state phase transitions}
\author[]{Nengji Zhou\inst{1,2}}
\author[]{Yuyu Zhang\inst{3}}
\author[]{Zhiguo L\"u\inst{4}}
\author[]{Yang Zhao\inst{2}\footnote{Corresponding author\quad E-mail:~\textsf{YZhao@ntu.edu.sg}}}
\address[1]{Department of Physics, Hangzhou Normal University, Hangzhou 310046, China}
\address[2]{Division of Materials Science, Nanyang Technological University, Singapore 639798, Singapore}
\address[3]{Department of Physics, Chongqing University, Chongqing 401331, China}
\address[4]{Key Laboratory of Artificial Structures and Quantum Control (Ministry of Education), Department of Physics, Shanghai JiaoTong University, Shanghai 200240, China}
\shortauthors{N.J. Zhou et al.}
\begin{abstract}
{\color{red}By means of a trial wave function, the multi-D$_1$ ansatz, extensive variational calculations  with more than ten thousand parameters have been} carried out to study quantum phase transitions in the ground states of a two-impurity system embedded in a common Ohmic bath of bosons. Quantum criticality in both the impurity system and the Ohmic bosonic bath is investigated with relevant transition points and critical exponents determined accurately. {\color{red}With the linear grid of the Ohmic spectral density, our numerical calculations produce a much better description of the ground states with lower energies than other calculations employing a logarithmic grid with a discretization factor far greater than unity. It offers a possible solution to the considerable controversy on the critical coupling in the literature. Moreover, the ground-state phase transition is inferred to be of first order in the presence of strong antiferromagnetic spin-spin coupling}, at variance with that in the ferromagnetic regime or in the absence of spin-spin coupling where the transition belongs to the Kosterlitz-Thouless universality class.
\end{abstract}
\shortabstract
\begin{document}
\maketitle

\section{Introduction}
Considerable attention has been devoted to open quantum systems over the past decades \cite{leg87, bre07, hur10}. A variety of open quantum systems, together with associated quantum dynamics and phase transitions, have been the subjects of extensive numerical and analytical investigations in a wide range of settings, such as the defect tunneling in solids \cite{bra92,voj01}, spontaneous emission in quantum optics \cite{rza75, gar17}, charge transfer processes in biological reactions \cite{eng07,col09}, semiconducting quantum dots in nano-cavities \cite{val10,ota11}, and quantum tunneling in superconducting circuits \cite{did12, bak14, hua18}. The spin-boson model (SBM) describes a single two-level system, e.g., a spin-$1/2$ particle or a magnetic impurity, linear coupled to environmental degrees of freedom represented by a continuous bath consisting of bosonic field modes \cite{wei07}. In spite of its apparent simplicity, the system possesses multifaceted properties in statics, dynamics, and quantum criticality, and has often been used as a paradigmatic model for studying open quantum systems.

There exists a ground-state phase transition in SBM separating a nondegenerate delocalized phase from a doubly degenerate
localized phase, provided that the coupling between the system and the environment can be characterized by a gapless spectral function $J(\omega)= 2\alpha \omega_c^{1-s}\omega^s$, where $\alpha$ denotes the dimensionless coupling strength, $\omega_c$ represents the high frequency cutoff, and $s$ is the spectral exponent \cite{hur08}. The phase transition is of second order in the sub-Ohmic regime ($0<s < 1$), and of the Kosterlitz-Thouless type in the Ohmic case ($s = 1$). The latter can be realized in the context of waveguide quantum electrodynamics by coupling a superconducting qubit to a uniform Josephson junction array \cite{gol13,per13,sny15}. In recent years, the transition boundary and the critical exponents have been estimated by a variety of approaches, such as the numerical renormalization group (NRG), exact diagonalization (ED), variational matrix product states (VMPS), quantum Monte Carlo (QMC), and variational methods (VM) \cite{voj05,alv09,win09,zha10,chin11,guo12,zhe13}. {\color{red}Numerical results of the critical couplings agree well in the deep sub-Ohmic regime $s<0.5$, but differ considerably in the shallow one $s>0.5$, let alone those in the Ohmic case $s=1$ due to the sensitivity of Kosterlitz-Thouless transition.} Very recently, a variational approach based on a systematic coherent-state expansion has been used to probe the ground-state properties of SBM at $s=1$ \cite{ber14}.

Variants of the standard SBM have also been intensively studied \cite{mul04, guo12, zhou14, zhou15, wal16, liu17}. The two-impurity SBM, for example, is one of the prominent SBM generalizations, as it is essential for studying quantum logical operations involving two qubits in quantum computations \cite{bon13,den16,pin17}. In addition, the presence of a second impurity allows us to address the interplay between quantum control and dissipation represented by the spin-spin and spin-bath interactions, respectively, giving rise to a much richer phase diagram \cite{ort10}. For the single-impurity SBM with an Ohmic bath, the Kosterlitz-Thouless transition point of $\alpha_c \approx 1$ is generally accepted in the weak tunneling limit, which can be estimated by mapping onto the anisotropic Kondo model with bosonization techniques \cite{leg87, gui85,hur08}. In contrast, the value of $\alpha_c$ for the two-impurity SBM attached a common Ohmic bath is still under contention. Specifically, $\alpha_{\rm c}=0.5$ was predicted variationally in the absence of both bias and spin-spin coupling \cite{mcc10}, very different from $\alpha_c \approx 0.18$ obtained in NRG calculations, $0.22$ in QMC simulations, and $0.16$ from nonequilibrium quantum dynamics with a tunneling constant of $0.025$ \cite{ort10,win14, hen16}. Very recently, $\alpha_{\rm c}=0.125$ was arrived at by means of the variational treatment based on a new ansatz and mean-field approximation \cite{zhe15}. It follows that an accurate determination of the transition point is needed for the two-impurity model. More importantly, it should be addressed whether the transition belongs to the Kosterlitz-Thouless universality class in the presence of the impurity-impurity coupling.

In previous studies on quantum phase transitions in SBM, attention was mainly focused on the spin-related observations, especially for the spontaneous magnetization, a nature order parameter \cite{chin11, naz12,fre13,lv13, bru14}.
The Bethe Ansatz solution of the anisotropic Kondo model, believed to be equivalent to the Ohmic single-spin model, shows that the spin coherence $\langle \sigma_x \rangle$ decreases monotonically with the coupling $\alpha$, and remains continuous and finite at the Kosterlitz-Thouless transition.
In contrast, a discontinuous reduction in the spin magnetization occurs at the transition point $\alpha_c$ from $\langle \sigma_z\rangle =0$ in the delocalized phase to $\langle \sigma_z\rangle=-1$ (or $1$) in the localized phase \cite{hur08}. Even under a tiny bias, however, the discontinuity in the magnetization is replaced by a smooth crossover, which may result in a significant reduction of the critical coupling $\alpha_c$.
Those results have been numerically confirmed by applying the NRG and VM approaches to the spin-boson model \cite{naz12,ber14}.
For the two-impurity SBM, unfortunately, small but finite bias values, e.g., $10^{-8}\omega_c$ and $10^{-5}\omega_c$, were sometimes chosen to trigger the phase transition. Therefore, results such as $\alpha_{\rm c}\approx 0.18$ and $0.125$ are possibly underestimated \cite{ort10,zhe15}. Moreover, the abrupt jumps in the magnetization has not been reproduced numerically in the absence of the bias. Beyond the magnetization, the bath observables provide a direct observation of the quantum criticality intrinsic to the environment possessing many-body effects. The critical properties of the sub-Ohmic bath have recently been explored with the help of a variational approach in which the complete spin-environment wave function of the ground state can be determined \cite{blu17}. However, whether they are good indicators for phase transition detection in the Ohmic SBM remains an open issue.

In this paper, a numerical variational method (NVM) is devised to study the ground-state phase transition of the two-impurity SBM in a common Ohmic bath. Using the generalized trial wave function composed of coherent-state expansions, which has been proven successful in treating the ground-state phase transitions and quantum dynamics of quantum many-body systems \cite{zhou14, zhou15, zhou15b,wan16,wan17,fuj17}, we systematically investigate the ground-state energy, magnetization, and spin coherence as well as the observables related to the Ohmic bath. The transition point and critical exponents are accurately determined, and phase diagram spanned by the spin-spin and spin-bath couplings is identified, in comparison with those obtained from NRG and QMC. The rest of the paper is organized as follows. In Sec.~$2$, the two-impurity model and NVM is described. In Sec.~$3$, the numerical results are presented for the validity of NVM, quantum criticality of the Ohmic bath, and phase diagram. Finally, they are discussed at length before drawing conclusions in Sec.~4.

\section{Model and Method}
For completeness and further reference, we first introduce the standard Hamiltonian of SBM,
\begin{equation}
\label{Ohami}
\hat{H} =  \frac{\varepsilon}{2}\sigma_z-\frac{\Delta}{2}\sigma_x + \sum_{k} \omega_k b_{k}^\dag b_{k}  +  \frac{\sigma_z}{2}\sum_k \lambda_k(b^\dag_{k}+b_{k}),
\end{equation}
where $\varepsilon$ is an energy bias,  $\Delta$ denotes the bare tunneling amplitude, $\sigma_x$ and $\sigma_z$ represent the Pauli spin-$1/2$ operators,
$b^\dag_k$ ($b_k$) is the bosonic creation (annihilation) operator of the $k$-th bath mode whose frequency is $\omega_k$, and $\lambda_k$ signifies the coupling amplitude between the spin and environment. By dividing the phonon frequency domain $[0, \omega_c]$ into $M$ intervals $[\Lambda_k, \Lambda_{k+1}]\omega_c$ ($k=0, 1, \ldots, M-1$), we calculate the coupling strength $\lambda_k$ and bosonic frequency $\omega_k$ in Eq.~(\ref{Ohami}) with a coarse-grained treatment of the continuous spectral density function $J(\omega)=\sum_k\lambda_k^2\delta(\omega-\omega_k)$ \cite{bul05, voj05,zha10,zhou14, blu17},
\begin{equation}
\label{sbm1_dis}
\lambda_k^2  =  \int^{\Lambda_{k+1}\omega_c}_{\Lambda_k\omega_c}dt J(t), \quad \omega_k  =  \lambda^{-2}_k \int^{\Lambda_{k+1}\omega_c}_{\Lambda_k\omega_c}dtJ(t)t.
\end{equation}
For convergence, we set the cutoff frequency $\omega_c=1$. A logarithmic discretization procedure with the parameter
$\Lambda_k=\Lambda^{k-M}$ is usually adopted for the sub-Ohmic regime ($s<1$) presenting a second-order phase transition. However,
in a more prominent case with Ohmic spectrum, i.e., $s=1$, the continuum limit $\Lambda \rightarrow 1$ is required to obtain an accurate quantum criticality
of the Kosterlitz-Thouless transition {\color{red}\cite{bul05,naz12,mcc10}}, while $\Lambda=1.4\sim2.0$ is used in earlier numerical works \cite{ort10,ber14}. A linear discretization procedure with $\Lambda_k=k/M$ is a possible alternative, for the bosonic modes at different frequencies are equally important. {\color{red}Hence, unless noted otherwise the NVM results presented here are obtained with the linear discretization.}

In this paper, we primarily aim to study the two-impurity SBM for which the Hamiltonian is given by
\begin{eqnarray}
\label{Ohami_twospin}
\hat{H} & = & \frac{\varepsilon}{2}(\sigma_1^z+\sigma_2^z)-\frac{\Delta}{2}(\sigma_1^x+\sigma_2^x) + \sum_{k} \omega_k b_{k}^\dag b_{k} \nonumber \\
& + & \frac{\sigma_1^z+\sigma_2^z}{2}\sum_k \lambda_k(b^\dag_{k}+b_{k}) + \frac{K}{4}\sigma_1^z\sigma_2^z,
\end{eqnarray}
where the subscripts of $\sigma_i~(i=1, 2)$ correspond to qubits $1$ and $2$, and $K$ is the Ising-type qubit-qubit interaction.
First of all, a simple case in the absence of bias and spin-spin interaction $K=\varepsilon=0$ is investigated
for understanding the ground-state quantum phase transition in the Ohmic environment. The transition point is  determined accurately, and critical properties of
the Ohmic bath are identified. After that, both the ferromagnetic ($K<0$) and  antiferromagnetic ($K>0$) situations as well as the biased cases ($\varepsilon >0$)
are studied, and the phase diagram is given in Subsection $3.3$.

As one of successful approaches that enables direct access to the ground-state wave function, the variational method has recently been adopted to study SBM, where the form of the trial wave function plays a vital role in obtaining the ground state \cite{naz12,zhou14,blu17}. In this work, a systematic coherent-state expansion, termed as the ``multi-D$_1$ ansatz'', is used as the variational ansatz, which has been proved to be efficient in tackling the ground-state phase transitions and quantum dynamics of SBM and its variant \cite{zhou14,zhou15,wan16},
{\color{red}
\begin{eqnarray}
\label{vmwave}
|\Psi \rangle & = & |\uparrow\uparrow \rangle |\rm{B}_{\uparrow\uparrow}\rangle +  |\uparrow\downarrow \rangle |\rm{B}_{\uparrow\downarrow}\rangle +  |\downarrow\uparrow \rangle |\rm{B}_{\downarrow\uparrow}\rangle +  |\downarrow\downarrow \rangle |\rm{B}_{\downarrow\downarrow}\rangle \nonumber \\
&=&|\uparrow\uparrow \rangle \sum_{n=1}^{N} A_n \exp\left[ \sum_{k=1}^{M}\left(f_{n,k}b_k^{\dag} - \mbox{H}.\mbox{c}.\right)\right] |0\rangle_{\textrm{b}} \nonumber \\
& + & |\uparrow\downarrow \rangle \sum_{n=1}^{N} B_n \exp\left[ \sum_{k=1}^{M}\left(g_{n,k}b_k^{\dag} - \mbox{H}.\mbox{c}.\right)\right]
|0\rangle_{\textrm{b}}   \\
& + & |\downarrow\uparrow \rangle \sum_{n=1}^{N} C_n \exp\left[ \sum_{k=1}^{M}\left(h_{n,k}b_k^{\dag} - \mbox{H}.\mbox{c}.\right)\right]
|0\rangle_{\textrm{b}}  \nonumber \\
& + & |\downarrow\downarrow \rangle \sum_{n=1}^{N} D_n \exp\left[ \sum_{k=1}^{M}\left(p_{n,k}b_k^{\dag} - \mbox{H}.\mbox{c}.\right)\right]
|0\rangle_{\textrm{b}}, \nonumber
\end{eqnarray}
where $|\uparrow\uparrow \rangle |\rm{B}_{\uparrow\uparrow}\rangle$ represents one of the bases in the ansatz,} H$.$c$.$ denotes Hermitian conjugate, $\uparrow$ ($\downarrow$) stands for the spin up (down) state, and $|0\rangle_{\rm b}$ is the vacuum state of the bosonic bath. The variational parameters $f_{n,k},~ g_{n,k},~ h_{n,k}$, and $p_{n,k}$ represent the displacements of the coherent states  correlated to the spin configurations $|\uparrow\uparrow \rangle, |\uparrow\downarrow \rangle, |\downarrow\uparrow \rangle$, and $|\downarrow\downarrow \rangle$ respectively,  and $A_n,~ B_n,~ C_n$, and $D_n$ are weights of the coherent states. The subscripts $n$ and $k$ correspond to the ranks of the coherent superposition and effective bath mode, respectively.

Recently, it has been reported that the variational treatment based on Silbey-Harris ansatz fails for the ground-state phase transition of SBM,
since the imposed constraints $f_{n,k}=-p_{n,k}$ and $A_n=D_n$ are broken in the localized phase or the biased case \cite{rob84,chin11, naz12}. Here the multi-D$_1$ ansatz goes beyond that of Silbey and Harris. Moreover, the number of the flexible variational parameters is much larger than that in Silbey-Harris one and its recent extension \cite{zhou15}. For example, it has more than $12~000$ variational parameters if $M=500$ and $N=6$. The sophistication of the trail wave function ensures it is possible to obtain an accurate description of the ground state around the transition point, for the existence of huge quantum fluctuations and entanglements in the environment, though the variational procedure becomes quite difficult.

Numerical variational method (NVM) is then utilized to search for the ground state by minimizing the system energy $E$ with respect to variational parameters. With the multi-D$_1$ ansatz defined in Eq.~(\ref{vmwave}) at hand, $E=\mathcal{H}/\mathcal{N}$ can be calculated with the Hamiltonian expectation $\mathcal{H}=\langle \psi|\hat{H}|\psi\rangle$ and norm of the wave function $\mathcal{N}=\langle \psi |\psi\rangle$. A set of self-consistency equations are then derived
\begin{equation}
\label{vmit}
\frac{\partial \mathcal{H}}{\partial x_{i}} - E\frac{\partial \mathcal{N}}{\partial x_{i}} = 0,
\end{equation}
where $x_i~(i=1,~2,\cdots,~4NM+4N)$ denotes any variational parameter. For each set of the model parameters, more than $100$ initial states are used with different ($A_n,~ B_n,~ C_n$, and $D_n$) uniformly distributed within an interval $[-1, 1]$, and ($f_{n,k},~ g_{n,k},~ h_{n,k}$, and $p_{n,k}$) obeying classical displacement law $\pm\lambda_k/\omega_k$. Meanwhile, the simulated annealing algorithm is employed in variational calculations in order to escape from metastable states.
The termination criterion of the iteration procedure is max$\{x_i^{*} -x_i\} < 1\times 10^{-10}$. Thus, one can obtain the ground-state solution $|\Psi_{\rm g}\rangle$ with the minimum energy $E_{\rm g}$. The reader is referred to {\color{red} Section 1 of Supporting Information for more details.}

Besides the ground-state energy $E_g$ as well as spin coherence and magnetization $\langle \sigma_{x,z} \rangle=\langle \Psi_{\rm g}|\sigma_{x,z}|\Psi_{\rm g} \rangle$, the observables related to the Ohmic bath defined in {\color{red} Section 2 of Supporting Information} are also evaluated as indicators to characterize ground-state phase transition for the two-impurity model, including the variances of phase space variables $\Delta X_{\rm b}$ and $\Delta P_{\rm b}$, the correlation functions $\rm Cor_X$ and $\rm Cor_P$, renormalized tunneling $\Delta_r$, and average displacements and coherent-state weights $\bar{f}_k, ~\bar{p}_k, ~\bar{A}$, and $\bar{D}$ \cite{ber14,blu17}.

Finally, convergence tests of variational results are performed against the numbers of effective bath modes $M$ and coherent-superposition states $N$. A linear discretization procedure is used  in this work for the Ohmic spectrum with $s=1$. {\color{red}The results in Section 3 of Supporting Information show that $M=500$ and $N=6$ are sufficient in NVM to rebuild the ground state of the two-impurity model.} Therefore, main results are presented with this setting unless noted otherwise.

\section{Numerical results}
\subsection{Validity of variational calculations}

\begin{figure}[tbp]
\centering
\subfigure[]{
\begin{minipage}[b]{0.35\textwidth}
\includegraphics[scale=0.35]{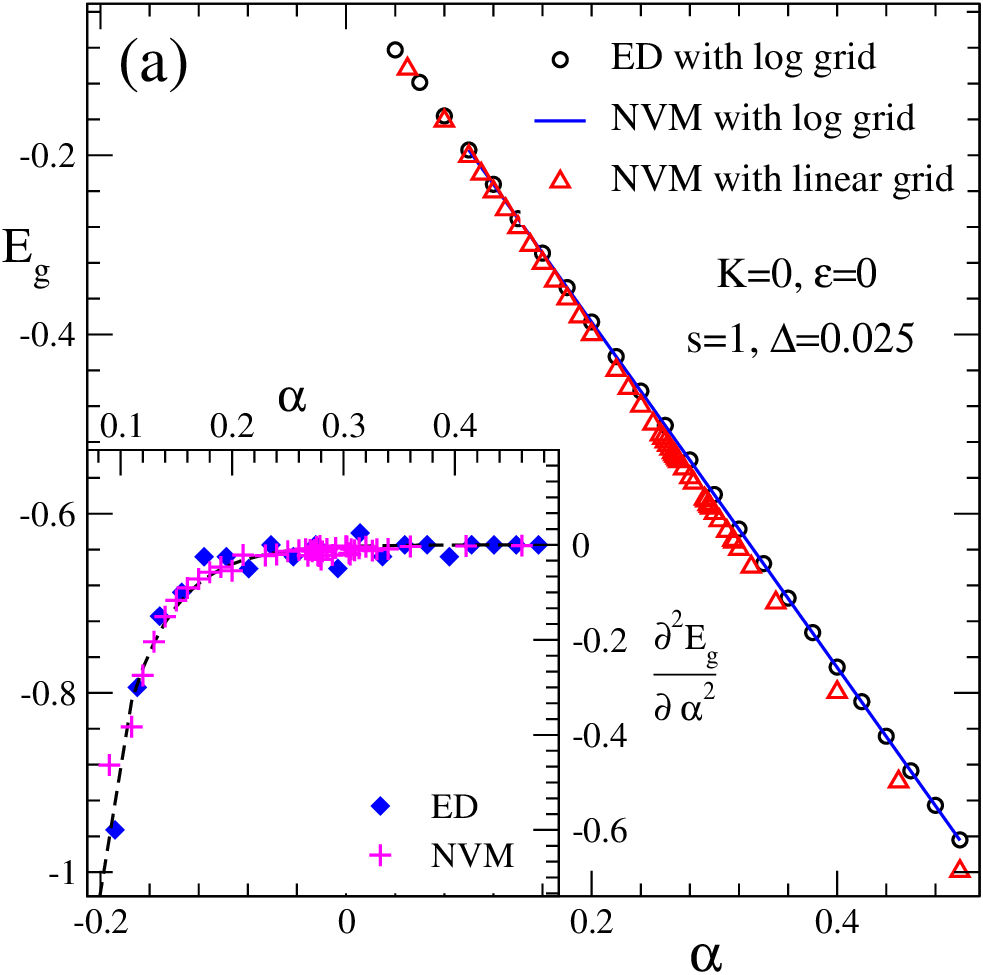}	
\end{minipage}
}
\subfigure[]{
\begin{minipage}[b]{0.35\textwidth}
\vspace{1.5\baselineskip}	
\includegraphics[scale=0.35]{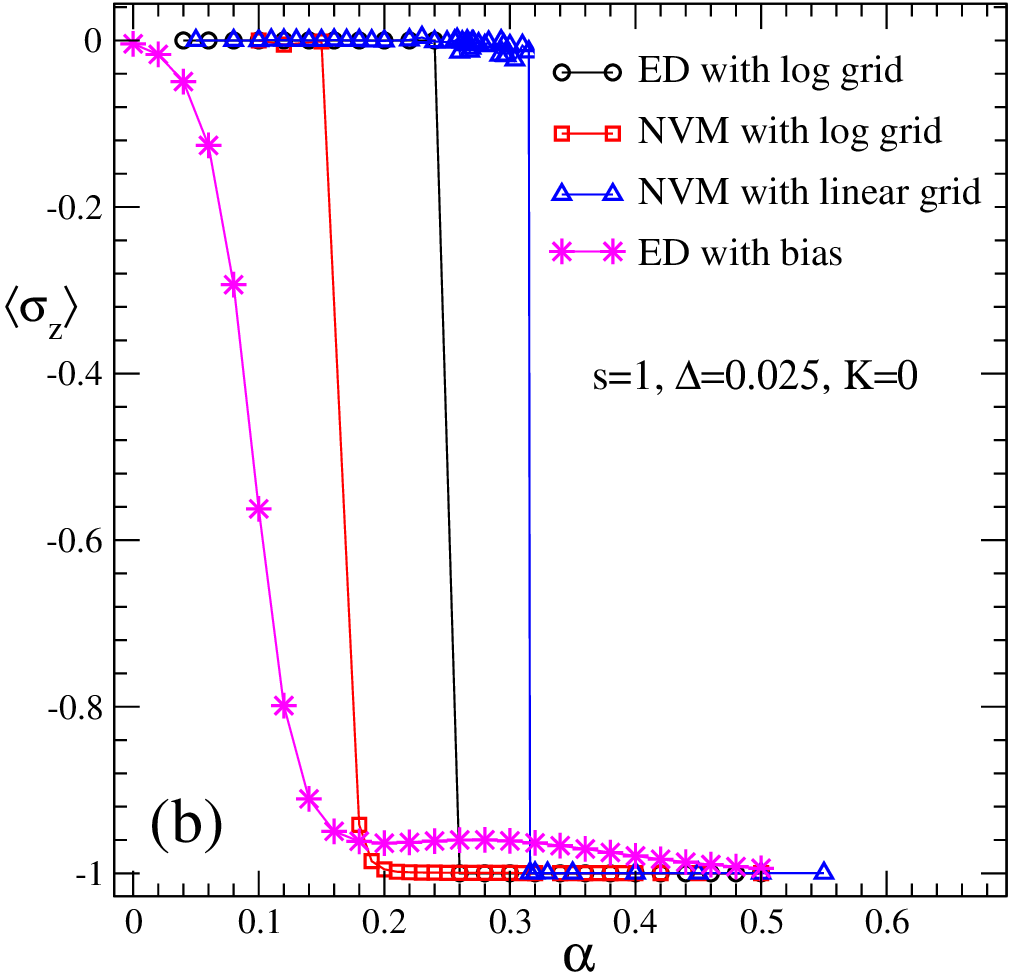}		
\end{minipage}
}
\vspace{-2.5\baselineskip}
\caption{ {\color{red}The ground-state energy $E_{\rm g}$ in (a) and magnetization $\langle \sigma_z\rangle$ in (b) of the Ohmic two-impurity SBM with $s=1$ as a function of the coupling strength $\alpha$ at $\varepsilon=K=0$ and $\Delta=0.025$. The numbers of effective bath modes and coherent-superposition states ($M=500, N=6$) and ($M=30, N=12$) are used for NVM with linear and logarithmic grids ($\Lambda=2$), respectively. The truncated number $N_{\rm tr} = 4$ is set in the exact-diagonalization (ED) procedure. In the inset of (a), the second derivative of $E_{\rm g}$ is shown for ED and NVM, and an exponential fit is presented with the dashed line. In (b), a bias case of $\epsilon=10^{-4}$ is given with stars.}
}
\label{f1}
\end{figure}

\begin{figure}[tbp]
\centering
\subfigure[]{
\begin{minipage}[b]{0.35\textwidth}
\includegraphics[scale=0.35]{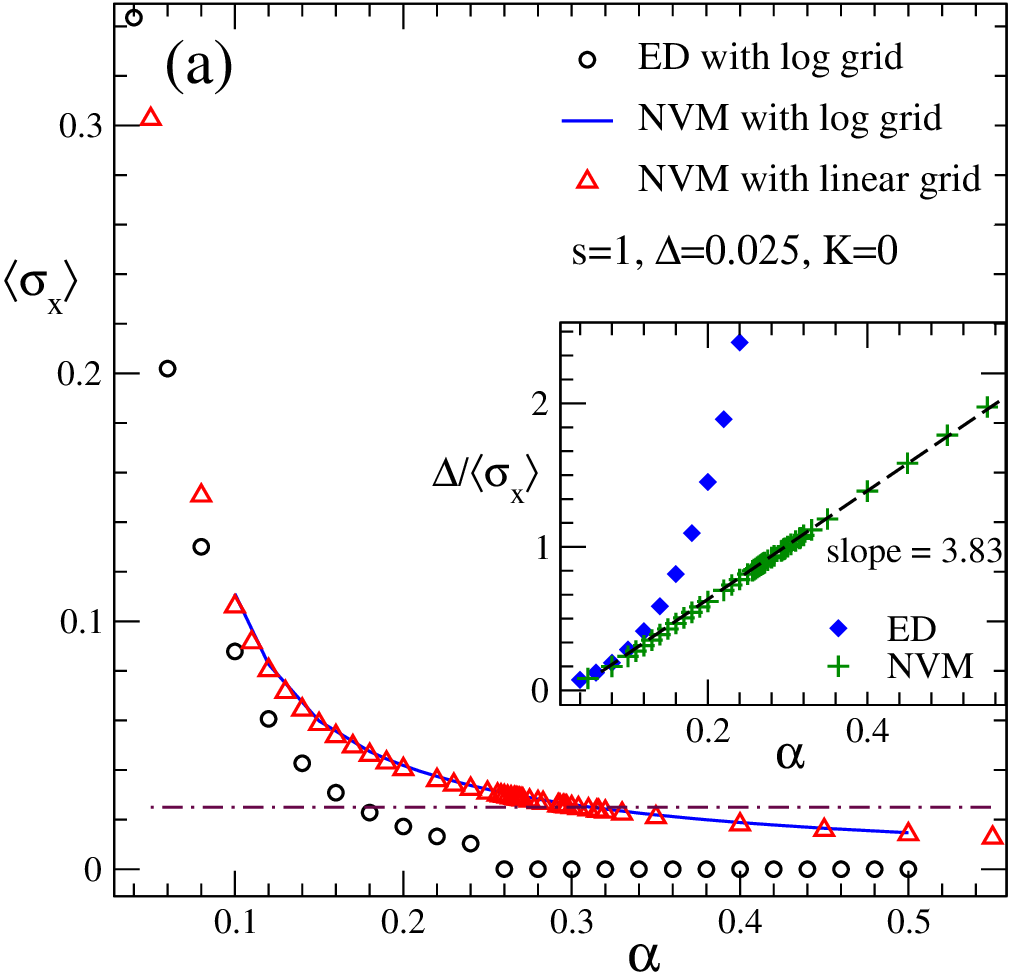}	
\end{minipage}
}
\subfigure[]{
\begin{minipage}[b]{0.35\textwidth}
\vspace{1.5\baselineskip}
\includegraphics[scale=0.35]{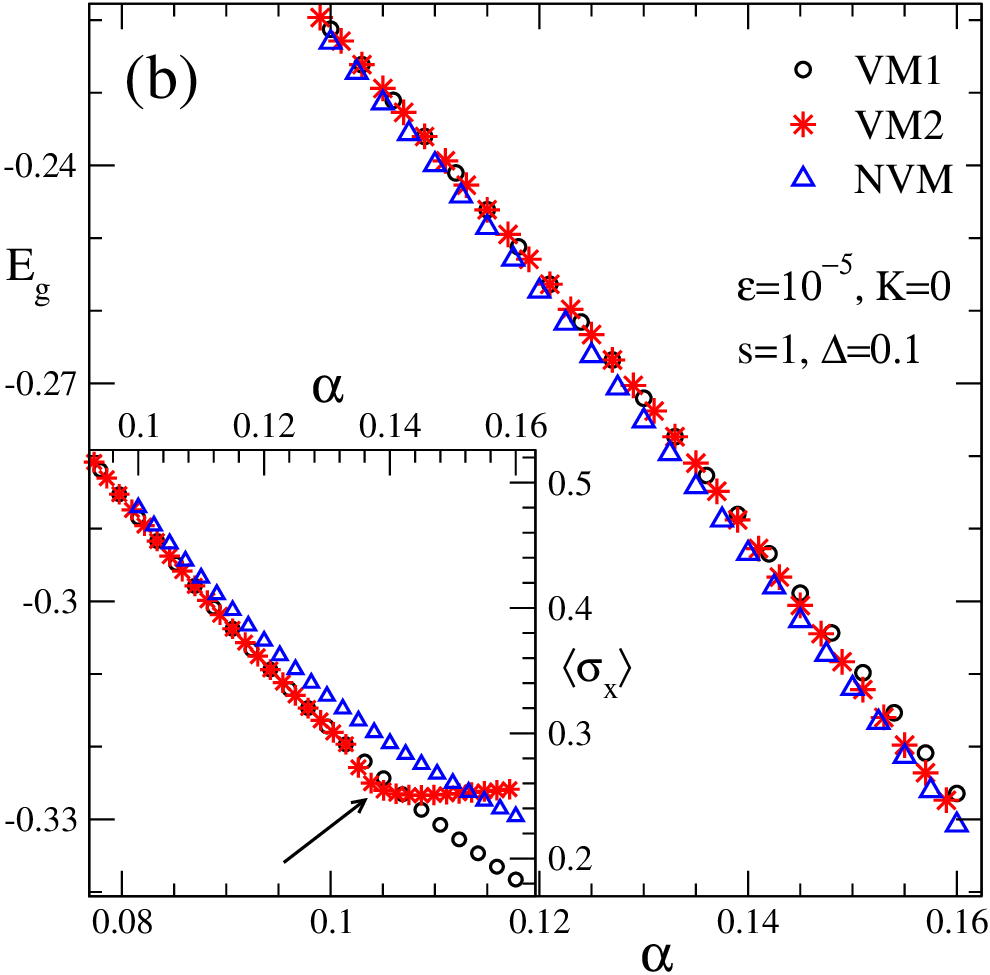}	
\end{minipage}
}
\vspace{-2.5\baselineskip}
\caption{{\color{red}(a) The spin coherence $\langle \sigma_x\rangle$ obtained by ED and NVM against the dissipation $\alpha$ at $\varepsilon=K=0,~s=1$, and $\Delta=0.025$ for the linear and logarithmic discretization ($\Lambda=2$). The criterion $\langle \sigma_x \rangle = \Delta/\omega_c$ is plotted with the dash-dotted line, and $\Delta / \langle \sigma_x\rangle $ is shown in the inset. (b) The ground-state energy $E_{\rm g}$ and spin coherence $\langle \sigma_x\rangle$ obtained from different variational works, i.e., $\rm VM1$ in Ref.\cite{mcc10}, $\rm VM2$ in Ref.\cite{zhe15}, and NVM in this work, in the biased case of $\varepsilon=10^{-5},~ s=1,~K=0$, and $\Delta=0.1$. A sharp kink of spin coherence is marked by the arrow.}
}
\label{f2}
\end{figure}

Using large-scale NVM simulations, we study the ground-state phase transition of the two-impurity SBM in the Ohmic regime, and compare our results with those from exact diagonalization (ED) \cite{zha10,zhou14}. As the ED technique is CPU-time and memory consuming,
we adopt the logarithmic discretization (instead of the linear one) in the ED procedure, together with a discretization
factor of $\Lambda=2$. {\color{red} Considering the constraint of available computational resources, an effective bath-mode number of $M = 12$ and a bosonic truncated number of $N_{\rm tr}=4$
are used here}. Without loss of generality, we focus on the case of $K=\varepsilon=0,~\Delta=0.025$, and $s=1$.

Fig.~\ref{f1}(a) shows the ground-state energy $E_{\rm g}$ as a function of the coupling strength $\alpha$.
{\color{red}For the $\alpha$ values considered here, the ground-state energies of NVM and ED are almost the same with a logarithmic grid, but
slightly higher than that of NVM with a linear grid. It is concluded that in the Ohmic regime the linear discretization yields a better approximation
to the ground state with lower energy than the logarithmic discretization with $\Lambda=2$. It is our belief that both NVM and ED approaches become exact
in the continuum limit $\Lambda \rightarrow 1$, though the latter cannot be achieved with computational resources currently available.}
In the inset of Fig.~\ref{f1}(a), the second derivative of $E_{\rm g}$ is plotted for further comparison. There is agreement between solid squares (ED) and pluses (NVM), and we obtain an exponential behavior of $\partial^2 E_{\rm g}/\partial \alpha^2$ by fitting the data to $y=a\exp(-bx)$, yielding a decay exponent of $b=23.2(2)$ {\color{red} and a root mean square error (RMSE) of about $0.01$ ($0.02$) for NVM (ED).} This lends support to a quantum phase transition of the Kosterlitz-Thouless type
for the two-impurity SBM in the presence of an Ohmic bath, as there is no discontinuity in derivatives of $E_{\rm g}$ of any order. {\color{red}By the discontinuity we mean the size of the jump far exceeds RMSE. }

In addition, magnetization $\langle \sigma_z \rangle$ as a function of $\alpha$ calculated by ED (NVM) is depicted in Fig.~\ref{f1}(b) with circles {\color{red}(squares for log grid and triangles for linear grid)}. For simplicity, only one branch of the doubly degenerate ground states is presented for $\langle \sigma_z \rangle\leq 0$,
and the other can be obtained easily by projecting the operator ${\cal P}_{x}=\sigma_x \exp(i\pi\sum b^\dag_{k}b_{k})$ onto $|\Psi_{\rm g}\rangle$.
In these curves, abrupt jumps from $\langle \sigma_z \rangle=0$ to $-1$ occur exactly at the Kosterlitz-Thouless transition, same as in the single-spin SBM \cite{hur08}. {\color{red} The transition point $\alpha_{\rm c}\approx 0.32$ of NVM with linear grid is then located by the discontinuity, greater than $\alpha_{\rm c}\approx 0.26$ from ED and $0.20$ from NVM with both methods using a logarithmic discretization factor of $\Lambda=2$.} In a biased case, e.g., $\varepsilon=1 \times 10^{-4}$, the sharp transition is softened to a smooth crossover, consistent with the results in Refs.~\cite{naz12,ber14,mcc10}.

The spin coherence $\langle\sigma_x\rangle$ as a function of $\alpha$ is also investigated for ED and NVM in Fig.~\ref{f2}(a).
{\color{red} The NVM coherence for both the logarithmic and the linear grid exhibits a monotonic, smooth decrease, while that of ED manifests a strong suppression in the entire $\alpha$ range.} A sudden drop to zero is found for the ED coherence when $\alpha \geq \alpha_{\rm c}$, at variance with the prediction that the coherence should be a continuous function of $\alpha$ retaining a finite value $\Delta/\omega_c$ at the Kosterlitz-Thouless transition in the presence of an Ohmic bath \cite{hur08,ber14}. With this criterion, the NVM critical point of $\alpha_{\rm c}\approx 0.3$ is indicated by the intersection of the triangles and the dot-dashed line, close to that obtained from the NVM magnetization curve in Fig.~\ref{f1}(b). In the inset, $\Delta/\langle\sigma_x\rangle$ calculated by NVM exhibits linear behavior conforming to the expression of $\Delta/\langle\sigma_x\rangle=(2\alpha-1)\omega_c$ obtained via the Bethe Ansatz solution for the single-spin model \cite{hur08}, although our slope is $3.83(3)$.  In contrast, the ED result deviates from the linear relation substantially, {\color{red} suggesting that the convergence is not reached with a truncated number of $N_{\rm tr}=4$.}

To further confirm the validity of NVM, we compare our variational results based on the multi-D$_1$ ansatz with those from other variational studies, named as ``VM$1$" \cite{mcc10} and ``VM$2$" \cite{zhe15} for convenience, in which the Silbey-Harris ansatz and its extension were utilized in combination with unitary transformations with variational parameters. In Fig.~\ref{f2}(b), a biased case with $\varepsilon=1\times10^{-5},~K=0,~\Delta=0.1$, and $s=1$ is considered. Shown in Fig.~\ref{f2}(b) as functions of $\alpha$ are the ground-state energy $E_{\rm g}$ and the spin coherence $\langle \sigma_x \rangle$ calculated by VM$1$ (circles), VM$2$ (stars), and NVM (triangles). The NVM result has lower ground-state energy and larger spin coherence, implying that the ground state obtained by our ansatz is indeed the most accurate among the three. Moreover, the arrow in the inset indicates a kink in $\langle \sigma_x \rangle$ calculated by VM$2$, against the usual expectation that $\langle \sigma_x \rangle$ decays monotonically and smoothly around the Kosterlitz-Thouless transition \cite{ber14}. The increase of the spin coherence as $\alpha$ goes above $0.14$ is counterintuitive, since strong environment-induced dissipation will destroy the quantum entanglement that preserves the spin coherence.

\begin{figure}[tbp]
\centering
\includegraphics[scale=0.45]{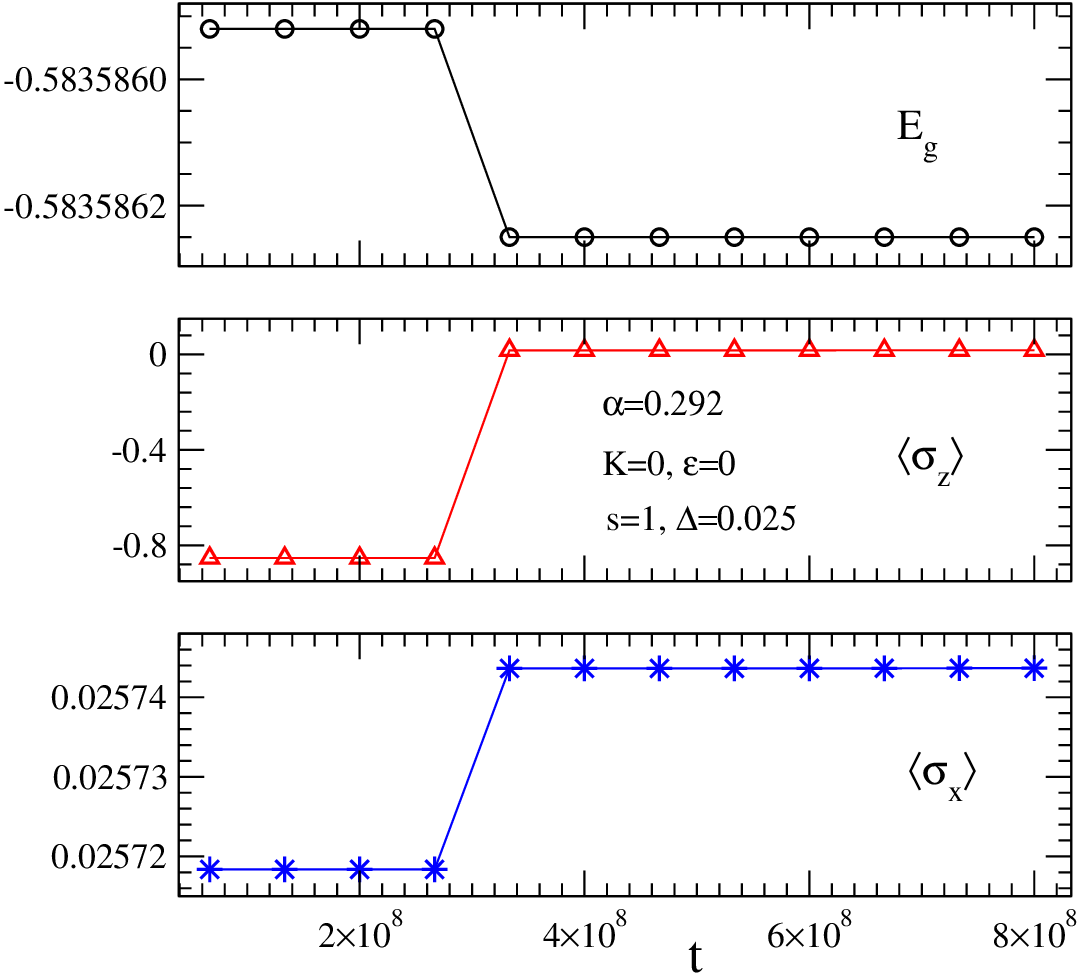}
\vspace{-1\baselineskip}
\caption{ Time evolution of the ground-state energy $E_{\rm g}$, magnetization $\langle \sigma_z\rangle$, and spin coherence $\langle \sigma_z\rangle$ for the coupling strength $\alpha=0.292$ at $\varepsilon=K=0,~ s=1$, and $\Delta=0.025$, where $t$ denotes the number of iterations.
}
\label{f3}
\end{figure}

In fact, numerous metastable states exist in the vicinity of the ground state around the Kosterlitz-Thouless transition. Taking $\alpha=0.292$ as an example, time evolution of $E_{\rm g},~\langle \sigma_z \rangle$, and $\langle \sigma_x \rangle$ is presented in Fig.~\ref{f3}. A huge change in the magnetization appears from $\langle \sigma_z \rangle=-0.85$ (a metastable state) to $0$ (the ground state), while the decrement in $E_{\rm g}$ (or the increment in $\langle\sigma_x\rangle$) is only a paltry amount of $3\times 10^{-7}$ ($3\times 10^{-5}$).
{\color{red} Usually, one can use the trapped time $\tau$ to characterize the metastability, which is related to the typical barrier height $\Delta E$ as $\ln \tau \sim \Delta E /kT$. Here the effective temperature $T$ depends upon the relaxation factor $t=0.1$ in our variational iteration procedure. Due to a complex energy landscape, there are many metastable states.} It follows that a certain high accuracy in computation is required to capture the ground state, beyond what can be afforded by the usual numerical methods in recent studies \cite{mcc10,ort10,win14,hen16,zhe15}. If a metastable state is mistaken as the ground state, the critical point is underestimated substantially, such as the biased ED results presented in Fig.~\ref{f1}(b). However, a non-zero bias value of $10^{-8}\omega_c$ is considered in previous NRG calculations {\color{red}\cite{ort10}}. To the best of our knowledge, the metastable states of the two-impurity models near the Kosterlitz-Thouless transition are discussed here for the first time, made possible only by the accuracy of the ground state obtained by our numerical calculations. In the iterative process of NVM in this work, the system always gets trapped in metastable states in the $\alpha$ range of $[0.28,~0.35]$.
To eliminate the inaccuracy due to metastable states near the quantum transition, additional hundreds of samples are then needed to achieve the true ground state, for each set of the model coefficients ($\alpha,~\Delta,~\varepsilon$, and $K$). In addition, some data from the metastates have to be discarded according to the criterion that the absolute value of $\langle \sigma_z \rangle$ monotonically increases with $\alpha$, and the sharp jump in $\langle \sigma_z \rangle$ is unique.

\begin{figure}[tbp]
\centering
\includegraphics[scale=0.45]{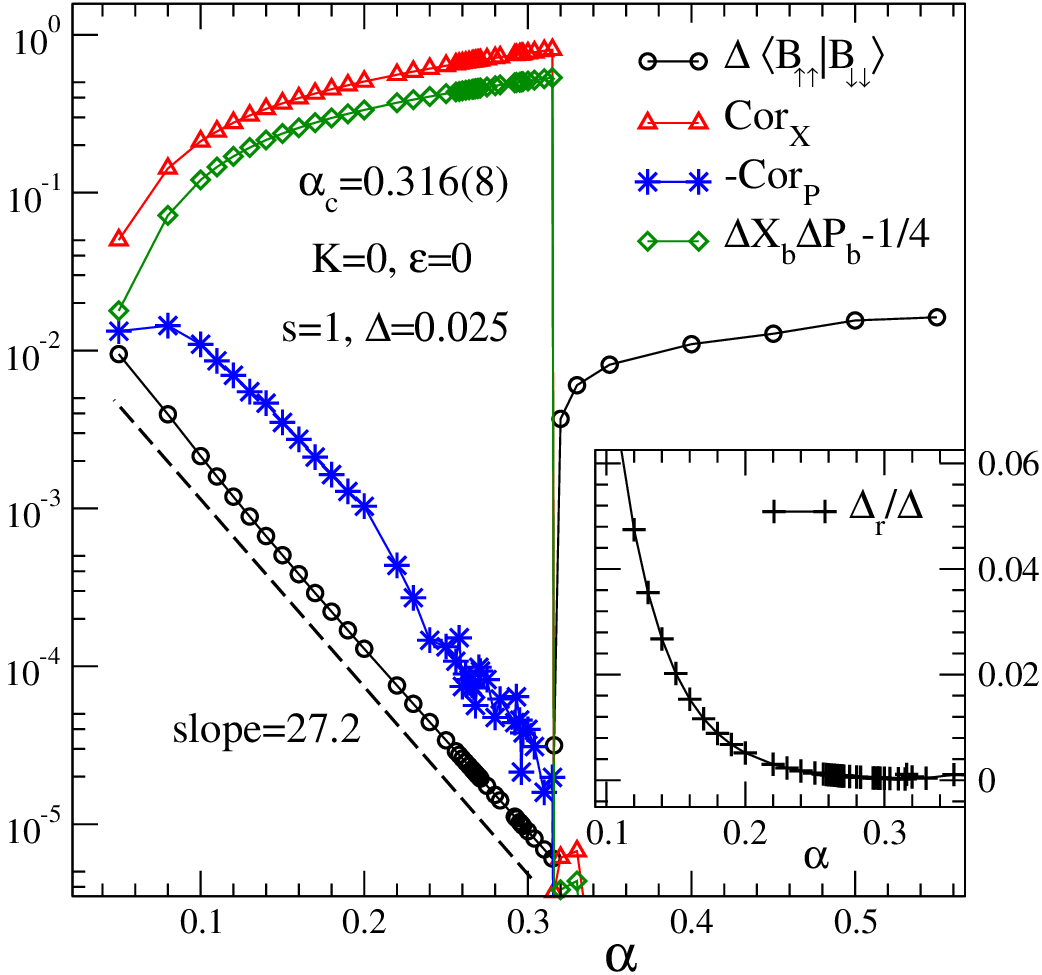}
\vspace{-1\baselineskip}
\caption{{\color{red}The correlation between two bases $\Delta \langle \rm{B}_{\uparrow\uparrow}|\rm{B}_{\downarrow\downarrow}\rangle$, correlation functions $\rm Cor_{X}$ and $\rm -Cor_{P}$ , and departure from the minimum uncertainty, $\Delta X_{\rm b}\Delta P_{\rm b}-1/4$ are shown with respect to the coupling strength $\alpha$. Dashed lines represent exponential fits. In the inset, the renormalized tunneling $\Delta_r/\Delta$ is given. }
}
\label{f4}
\end{figure}

\begin{figure}[tbp]
\centering
\subfigure[]{
\begin{minipage}[b]{0.35\textwidth}
\includegraphics[scale=0.35]{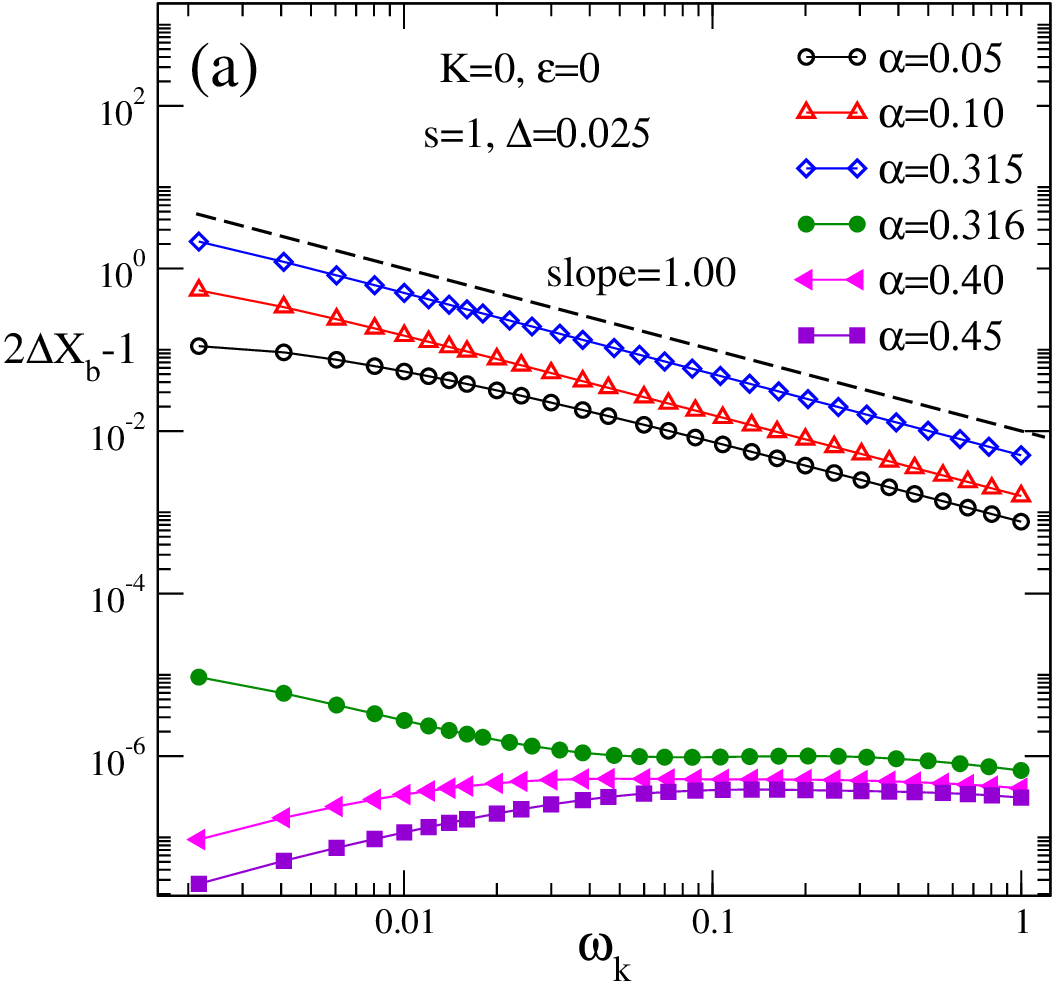}	
\end{minipage}
}
\subfigure[]{
\begin{minipage}[b]{0.35\textwidth}
\vspace{1.5\baselineskip}
\includegraphics[scale=0.35]{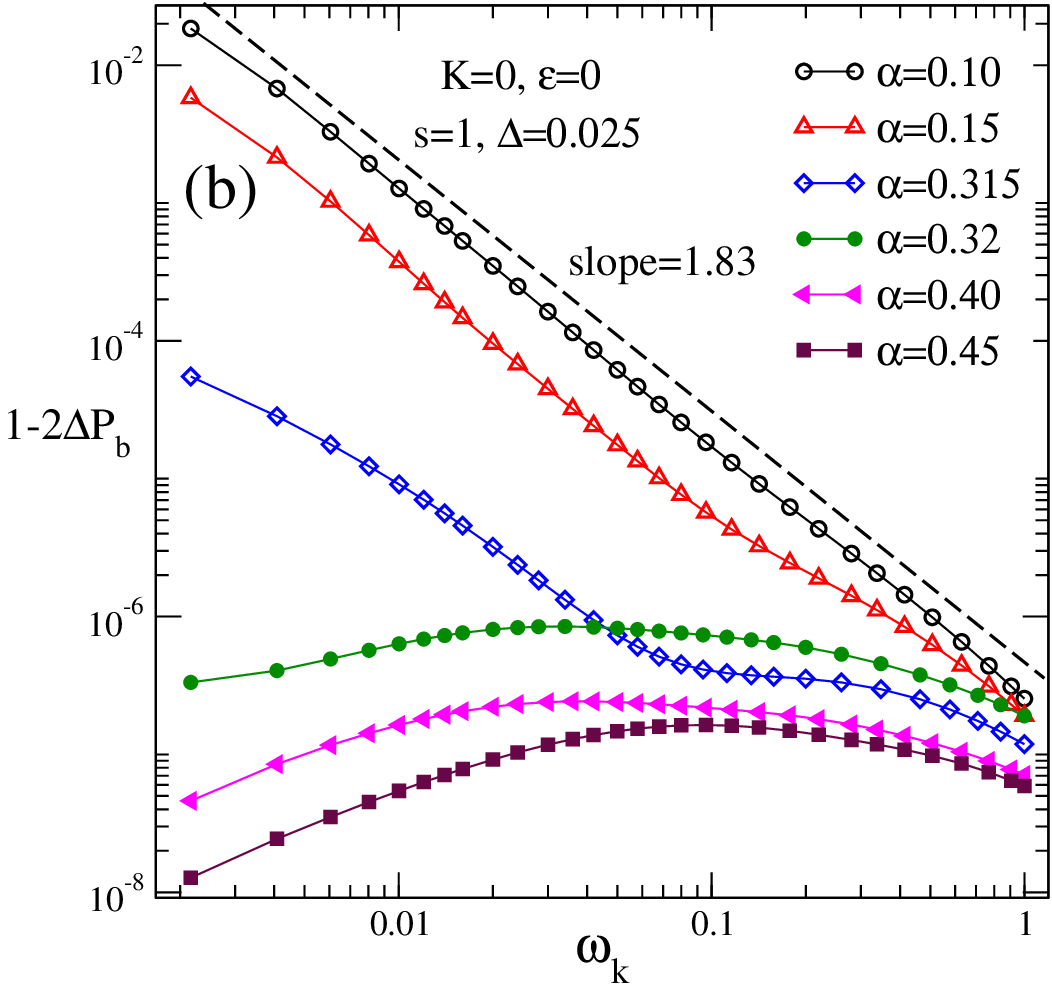}	
\end{minipage}
}
\vspace{-2.5\baselineskip}
\caption{ The variances of the phase space variables deviating from the equal uncertainty $\Delta X_{\rm b}=\Delta P_{\rm b}=1/2$ for different coupling strengthes $\alpha$ on a log-log scale at $\varepsilon=K=0,~s=1$, and $\Delta=0.025$. The dashed lines show power-law fits.
}
\label{f5}
\end{figure}

\subsection{Quantum criticality of the Ohmic bath}

In this subsection, we systematically study the ground-state properties of the Ohmic bath for the Kosterlitz-Thouless transition. Fig.~\ref{f4} displays $\alpha$-dependent, bath-related observables, as defined {\color{red} in the Sec.~$2$ of Supporting Information, which include the correlation between two bases $\langle \rm{B}_{\uparrow\uparrow} |\rm{B}_{\downarrow\downarrow}\rangle$}, renormalized tunneling $\Delta_{\rm r}$, the correlation functions $\rm Cor_{X}$ and $\rm -Cor_{P}$, and a measure of deviation from the uncertainty minimum, $\Delta X_{\rm b}\Delta P_{\rm b} - 1/4$. Note that for clarity only the bath mode with the lowest frequency $\omega_{k=0}$ is considered here for the variance of the phase space variables, and two bath modes with $l=0$ and $k=1$ for the correlation functions. In the delocalized phase (small $\alpha$), {\color{red}the correlation $\langle \rm{B}_{\uparrow\uparrow} |\rm{B}_{\downarrow\downarrow}\rangle$ scaled by the factor $\Delta$} quickly decays as $\alpha$ increases, and the curve is fitted using the dashed line of $\langle \rm{B}_{\uparrow\uparrow} |\rm{B}_{\downarrow\downarrow}\rangle \propto \exp(-27.2\alpha)$. {\color{red}In the localized phase (large $\alpha$), however, it gradually increases. As the basis correlation is the main component of the coherence correlation $\langle\sigma^{1}_x\sigma^{2}_x\rangle$, the nonzero value of $\langle \rm{B}_{\uparrow\uparrow} |\rm{B}_{\downarrow\downarrow}\rangle$ suggests that these two spins have strong correlation in the localized phase under the influence of the common Ohmic bath, though $\langle\sigma^{1}_x\rangle$ and $\langle\sigma^{2}_x\rangle$ vanish.} The transition point {\color{red}$\alpha_{\rm c}=0.316(8)$} is then determined according to the sudden jump of {\color{red}$\langle \rm{B}_{\uparrow\uparrow} |\rm{B}_{\downarrow\downarrow}\rangle$}. An exponential damping of $-\rm Cor_P$ is found for $\alpha < \alpha_c$ with the slope being comparable with that of $\langle \rm{B}_{\uparrow\uparrow} |\rm{B}_{\downarrow\downarrow}\rangle$. In contrast, there are increases in the correlation function $\rm Cor_X$ and the deviation from the uncertainty minimum, $\Delta X_{\rm b}\Delta P_{\rm b} - 1/4$. For $\alpha > \alpha_c$, however, the correlation functions $\rm Cor_P$ and $\rm Cor_X$ vanish, and so does the departure of the uncertainty from $1/4$. It is inferred that the bath modes in the localized phase are independent, and behave as a single-coherent state where $\Delta X_{\rm b}=\Delta P_{\rm b}=1/2$. {\color{red} In the inset, the decay of $\Delta_r/\Delta$ is displayed against the coupling $\alpha$, approaching to zero around the transition point $\alpha_c \approx 0.3$, consistent with the usual expectations.}

\begin{figure}[tbp]
\centering
\subfigure[]{
\begin{minipage}[b]{0.35\textwidth}
\includegraphics[scale=0.35]{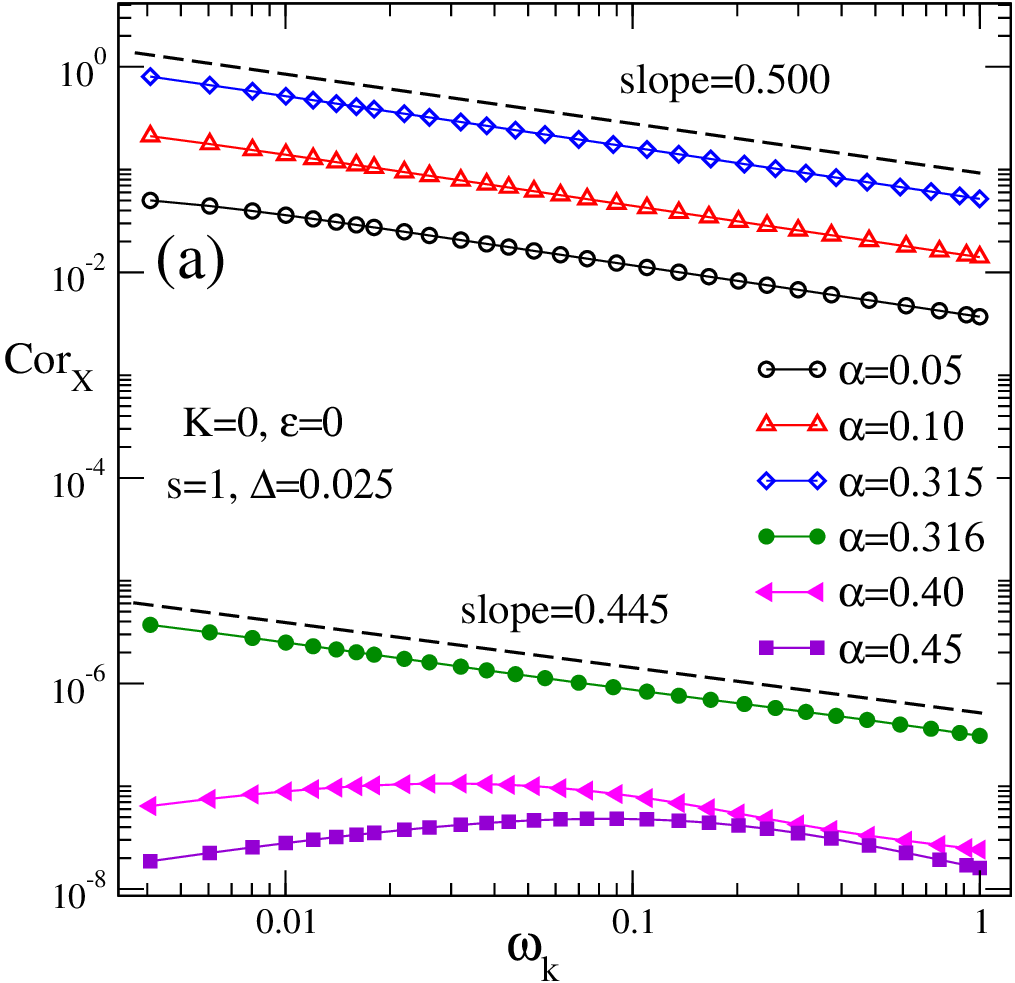}	
\end{minipage}
}
\subfigure[]{
\begin{minipage}[b]{0.35\textwidth}
\vspace{1.5\baselineskip}
\includegraphics[scale=0.35]{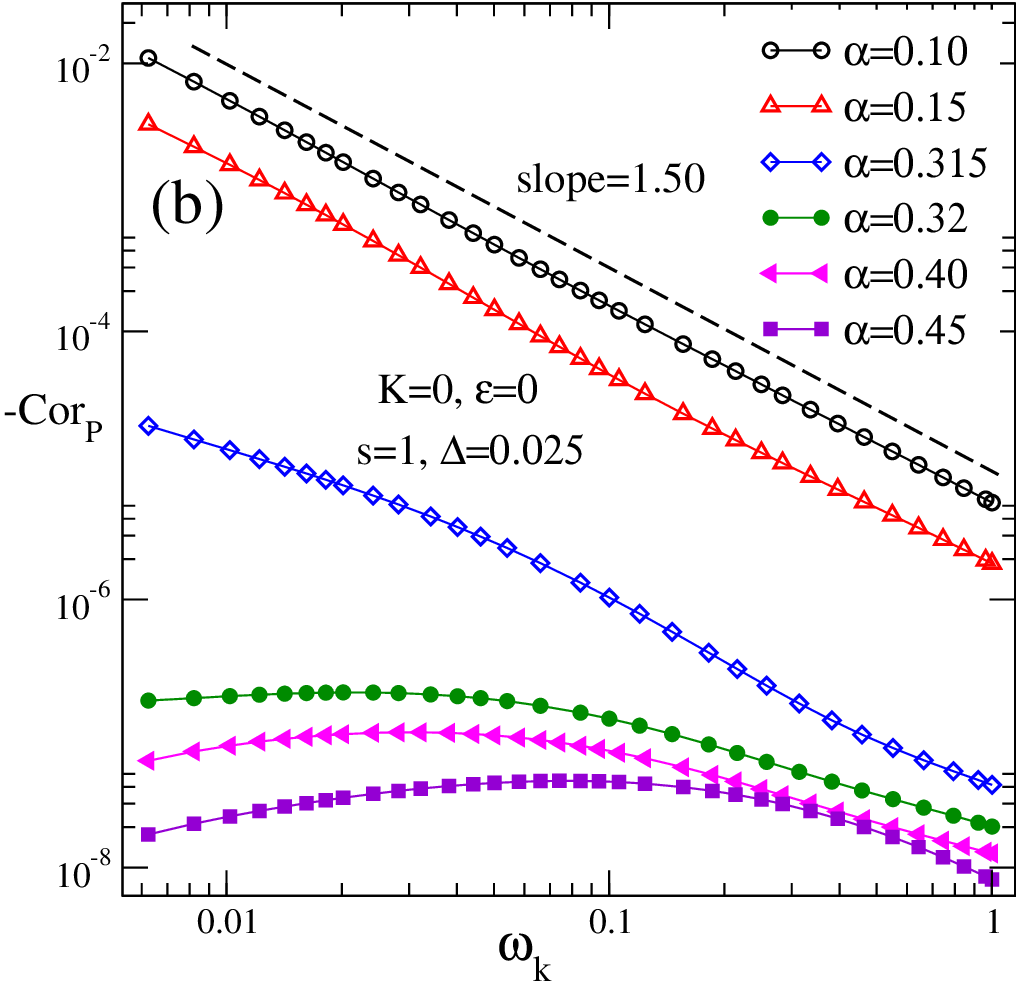}	
\end{minipage}
}
\vspace{-2.5\baselineskip}
\caption{ The correlation functions $\rm Cor_{X}$ in (a) and $\rm -Cor_{P}$ in (b) between two bath modes as a function of the bosonic frequency $\omega_k$ for different values of $\alpha$ on a log-log scale. Other parameters $\varepsilon=K=0,~ s=1$, and $\Delta=0.025$ are set. And dashed lines represent power-law fits.
}
\label{f6}
\end{figure}

In Fig.~\ref{f5}, the offsets $2\Delta X_{\rm b}-1$ and $1-2\Delta P_{\rm b}$ are plotted with respect to the frequency of the bath mode $\omega_k$ for various couplings $\alpha$.
In Fig.~\ref{f5}(a), perfect power-law decay is found in $2\Delta X_{\rm b}-1$ for $\alpha=0.315$ (top blue curve) with a critical exponent of $1.00(1)$.
A scaling property of $\Delta X_{\rm b}=0.0025/\omega_k+1/2$ is then uncovered with a power-law fit.
For cases of weaker coupling, such as $\alpha=0.05$ and $0.1$, almost parallel curves are found slightly below (and qualitatively similar to) that of $\alpha=0.315$.
For coupling strengthes larger than {\color{red} $\alpha_c=0.316(8)$}, e.g., $\alpha=0.40$ and $0.50$, $2\Delta X_{\rm b} - 1$ becomes negligibly small, signaling $\Delta X_{\rm b} = 1/2$ in the localized phase. Similar power-law decay of $1-2\Delta P_{\rm b}$ is found in Fig.~\ref{f5}(b) for $\alpha < \alpha_c$,  where the top curve corresponds to $\alpha=0.10$. An exponent of $1.83(2)$ can be obtained from the slope. If $\alpha > \alpha_c$, such as $\alpha=0.32,0.40$, and $0.50$,
both $1-2\Delta P_{\rm b}$ and $2\Delta X_{\rm b}-1$ increase with the frequency if $\omega_k < \omega^{\ast}$, followed by leveling off in the high-frequency regime (i.e., $\omega_k>\omega^{\ast}$) with a characteristic scale of $\omega^{\ast} \approx 0.1$. {\color{red} The similar nonmonotonic behavior was also reported for the ground-state transition
in the sub-Ohmic single-spin SBM \cite{blu17}.}

We next investigate the frequency dependence of the correlation functions $\rm Cor_{X}$ and $\rm -Cor_{P}$ defined in the Supporting material ``Sec. $2$ Observables related to the Ohmic bath'', where the lowest-frequency bosonic mode $l=0$ is fixed for convenience. As presented in Fig.~\ref{f6}, $\rm Cor_{X}$ and $\rm -Cor_{P}$ behave quite similarly with $\Delta X_{\rm b}$ and $1-2\Delta P_{\rm b}$, respectively, with exponents values of $0.500(3)$ and $1.50(2)$. It follows that $\rm Cor_X \sim 1/\sqrt{\omega_k}$ and $\rm Cor_P \sim - 1/\omega_k^{3/2}$ in the delocalized phase, while in the localized phase both of them become negligible. The vanishing value of the correlation function at any $\omega_k$ further supports that bath modes are independent of each other, despite being coupled to two impurities simultaneously. It seems that our multi-D$_1$ ansatz is capable to capture quantum entanglement properties built into the Ohmic bath in both the delocalized and localized phase.
\begin{figure}[tbp]
\centering
\subfigure[]{
\begin{minipage}[b]{0.35\textwidth}
\includegraphics[scale=0.35]{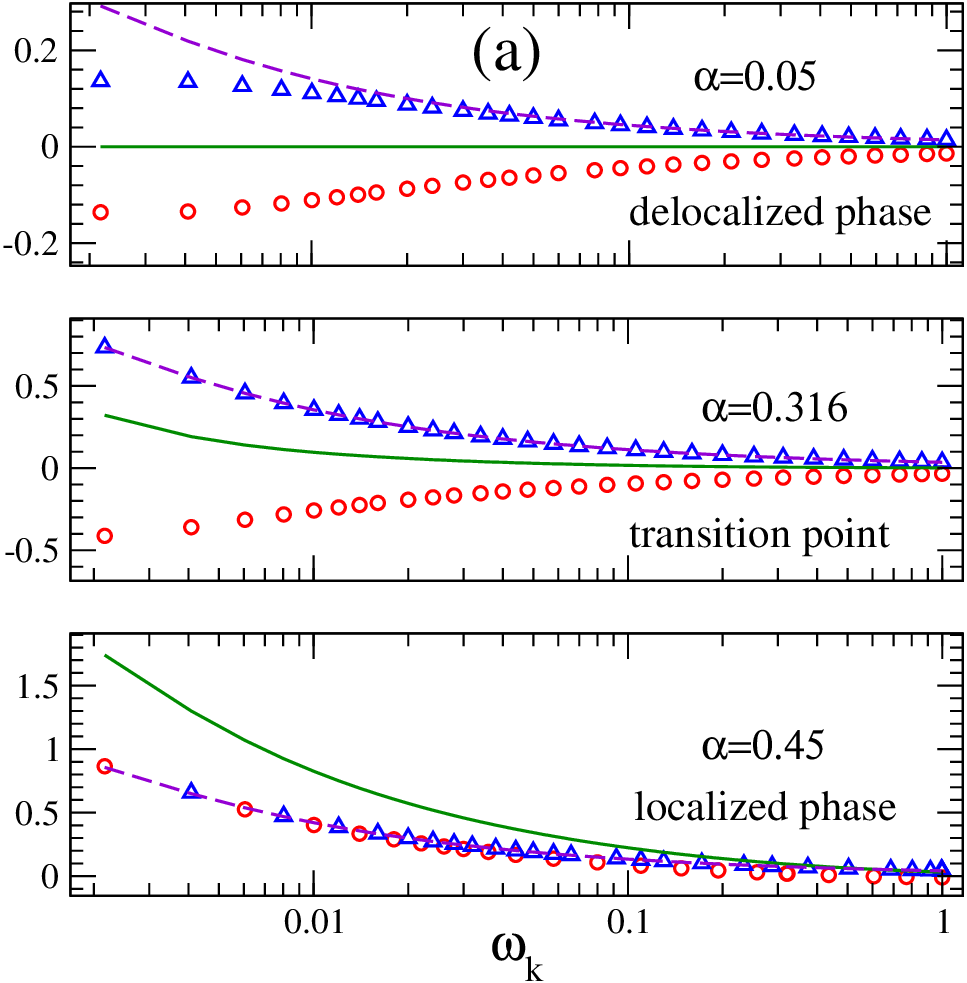}	
\end{minipage}
}
\subfigure[]{
\begin{minipage}[b]{0.35\textwidth}
\vspace{1.5\baselineskip}
\includegraphics[scale=0.35]{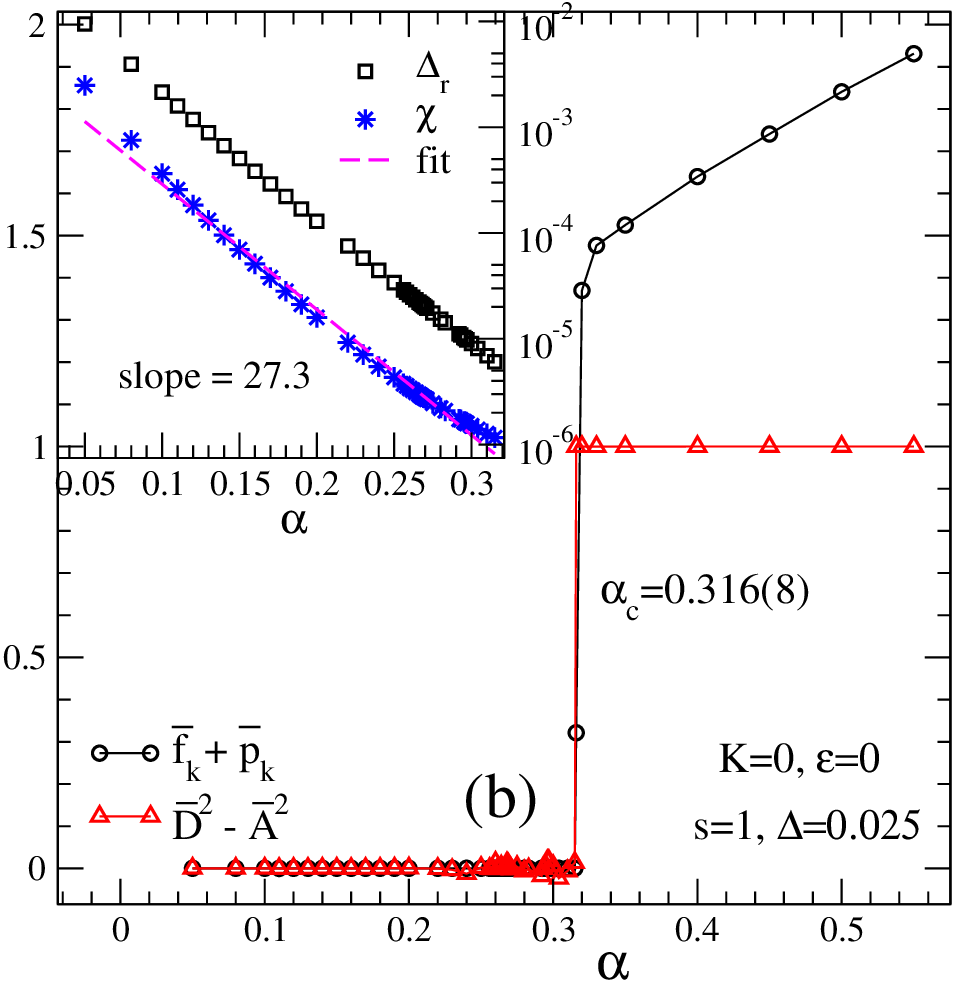}	
\end{minipage}
}
\vspace{-2.5\baselineskip}
\caption{ (a) Average displacement coefficients $\overline f_k$ (circles) and $\overline p_k$ (triangles) for three coupling strengthes $\alpha=0.05,~ 0.316$ and $0.45$, corresponding to the delocalized phase, transition point, and localized phase, respectively.  Solid lines represent the summations of $\overline f_k$ and $\overline p_k$, and dashed lines stand for the classical displacements $\lambda_k/\omega_k$. (b) The average displacements function $\overline f_k + \overline p_k$ and average weights function $\overline D^2-\overline A^2$ for the $0$-th bath mode. {\color{red} Inset shows the effective energy scale $\chi$ estimated from Eq.~(\ref{vm_chi}) and the renormalized tunneling $\Delta_r$ in the delocalized phase. The dashed line indicates an exponential fit.}
}
\label{f7}
\end{figure}

To better understand the critical properties of the Ohmic bath, we now turn our attention to the wave function of the ground state. The average coherent-state weights and bosonic displacements are calculated with Eqs.($21$-$22$) in the supporting file. Circles and triangles shown in Fig.~\ref{f7}(a) denote the displacement coefficients $\overline{f}_k$ and $\overline{p}_k$ respectively, for the coupling strengths $\alpha=0.05, 0.316$, and $0.45$ from top to bottom, corresponding to the delocalized phase, transition point, and localized phase, respectively. In the upper panel, the vanishing value of $\overline{f}_k+\overline{p}_k$ marked by the solid line manifests that the antisymmetry $\overline{f}_k=-\overline{p}_k$ develops naturally in the delocalized phase. At transition point, however, spontaneous symmetry breaking occurs in the low-frequency regime. The good coincidence of circles and triangles presented in the lower panel shows a symmetrical relation $\overline{f}_k=\overline{p}_k$ in the localized phase, contrary to the constrain condition imposed in Silbey-Harris ansatz \cite{rob84}. Focusing on the $0$-th bath mode, we demonstrate $\overline{f}_k+\overline{p}_k$ and $\overline{D}^2-\overline{A}^2$ as a function of $\alpha$ in Fig.~\ref{f7}(b). The critical point {\color{red} $\alpha_c=0.316(8)$} is again determined, the same as that obtained in Fig.~\ref{f4}.

For further comparison, the classical displacement $\pm\lambda_k/\omega_k$ estimated by the minimum of the static spin-dependent potential is also plotted in Fig.~\ref{f7}(a) with dashed lines. Distinguishable difference between the average displacement and classical one is found in the delocalized phase, but vanishes at the transition point and in the localized phase, in agreement with the general expectation \cite{mcc10}. Furthermore, an optimal displacement formula is proposed
\begin{equation}
\label{vm_chi}
|f_k|=|p_k|=\frac{\lambda_k}{\omega_k+\chi},
\end{equation}
where $\chi$ denotes an effective energy scale. {\color{red} Fitting our data to the above equation, $\chi$ is obtained as a function of $\alpha$, as presented in the inset of Fig.~\ref{f7}(b) with a slope of $27.3(3)$ for the exponential decay. For comparison, the renormalized tunneling $\Delta_r$ is also plotted at $\alpha < \alpha_c$. It is nearly parallel to $\chi$, supporting the usual assumption $\chi \propto \Delta_r$ \cite{naz12,ber14}.}

\subsection{Phase diagram}
\begin{figure}[tbp]
\centering
\subfigure[]{
\begin{minipage}[b]{0.35\textwidth}
\includegraphics[scale=0.35]{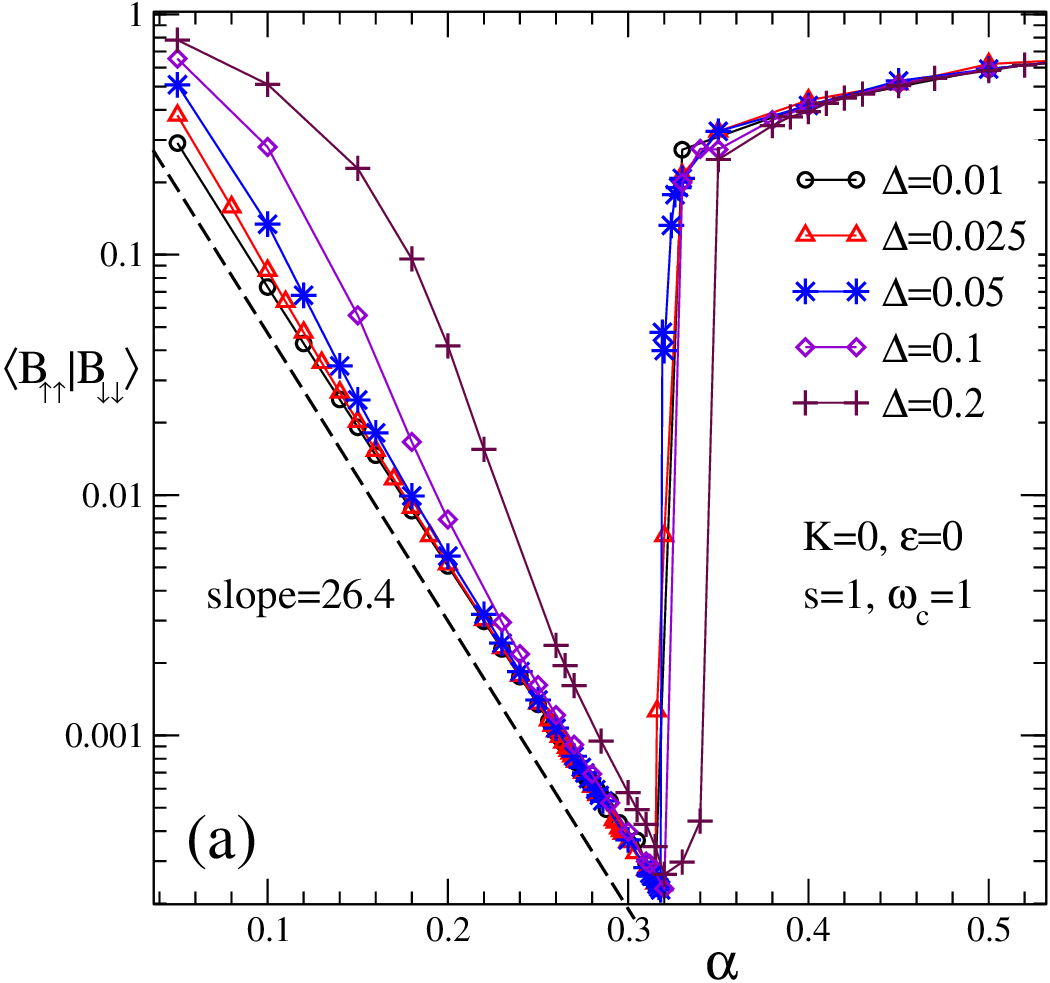}	
\end{minipage}
}
\subfigure[]{
\begin{minipage}[b]{0.35\textwidth}
\vspace{1.5\baselineskip}
\includegraphics[scale=0.35]{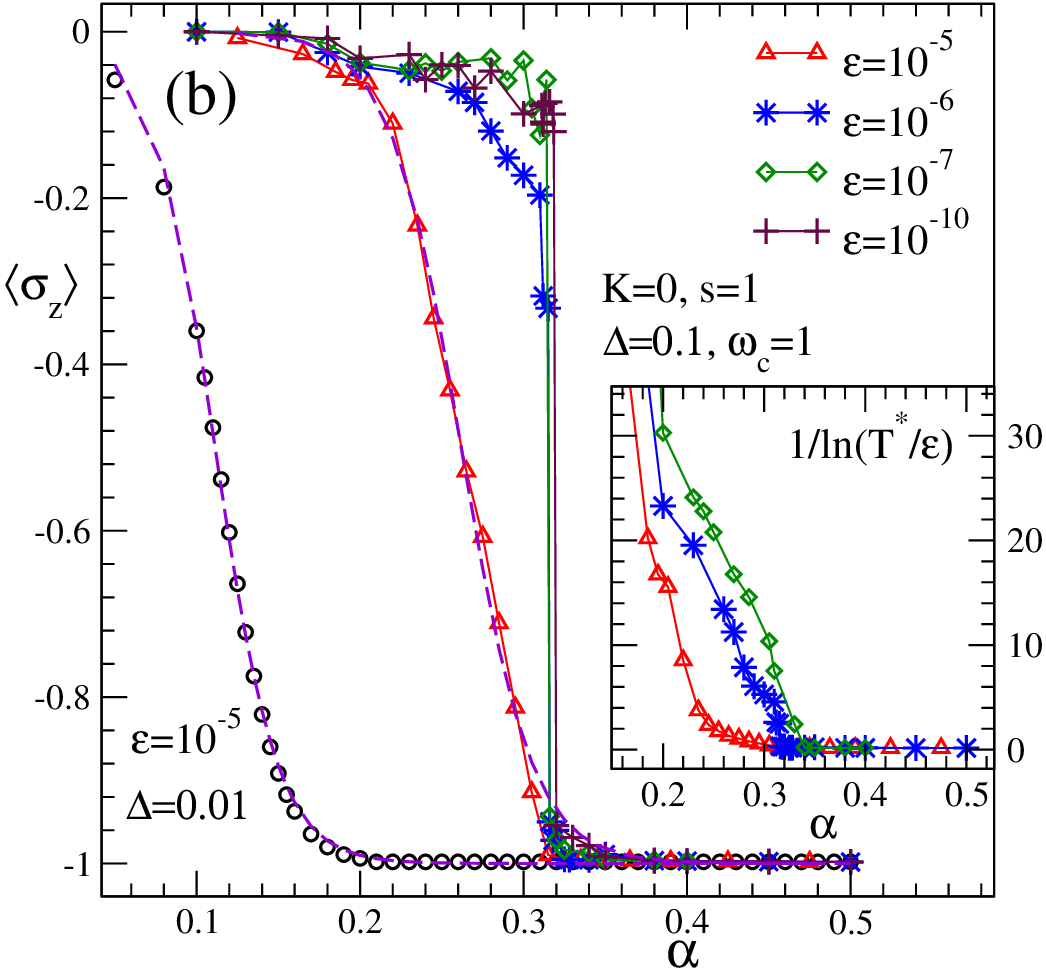}	
\end{minipage}
}
\vspace{-2.5\baselineskip}
\caption{ {\color{red}(a) The correlation between two bases $\langle \rm{B}_{\uparrow\uparrow}|\rm{B}_{\downarrow\downarrow}\rangle$ for different tunneling constants $\Delta=0.01,~ 0.025,~ 0.05,~ 0.1$, and $0.2$ at $\varepsilon=K=0,~ s=1$, and $\omega_c=1$.  The dashed line shows an exponential fit. (b) The magnetization $\langle \sigma_z\rangle$ as a function of $\alpha$ for different values of the bias $\varepsilon=10^{-5},~ 10^{-6},~ 10^{-7}$, and $10^{-10}$ at $\varepsilon=K=0,~ s=1$, and $\Delta=0.1$. In addition, the case of $\Delta=0.01$ is shown with circles, and exponential-like fits are presented with dashed lines. The scaling behavior of the crossover scale $T^{*}$ is shown in the inset.}
}
\label{f8}

\end{figure}
\begin{figure}[tbp]
\centering
\includegraphics[scale=0.45]{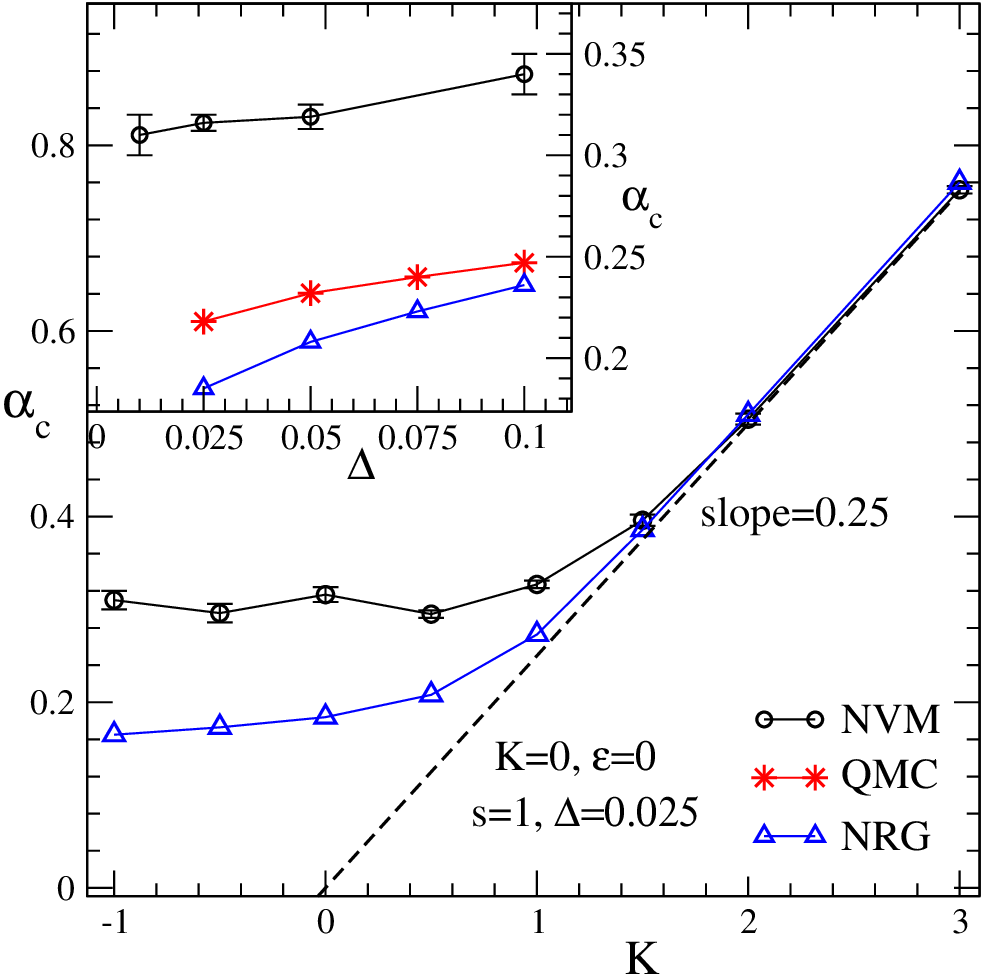}
\vspace{-1\baselineskip}
\caption{ Phase diagram of the Ohmic two-impurity SBM in the plane of the dissipation $\alpha$ and spin-spin coupling strength $K$. The dashed line represents the renormalized coupling $K_r=K-4\alpha\omega_c^s/s=0$. In the inset, transition boundary $\alpha_{\rm c}$ obtained from NVM is plotted as a function of the tunneling amplitude $\Delta$, in comparison with those of QMC and NRG in Refs.\cite{win14,ort10}.
}
\label{f9}
\end{figure}

Subsequently, we perform comprehensive NVM simulations to identify the influence of the tunneling constant $\Delta$, the bias field $\epsilon$, and the Ising coupling strength $K$ on the ground-state phase transition. {\color{red} The correlation between two bases $\langle \rm{B}_{\uparrow\uparrow}|\rm{B}_{\downarrow\downarrow}\rangle$} is plotted in Fig.~\ref{f8}(a) for various values of $\Delta$, and the transition points are determined. In a weak tunneling case of $\Delta=0.01$, an exponential decay is found in the delocalized phase with a slope of $26.4(3)$, compatible with that of $\Delta=0.025$ presented in Fig.~\ref{f4}. As $\Delta$ increases, {\color{red}$\langle \rm{B}_{\uparrow\uparrow}|\rm{B}_{\downarrow\downarrow}\rangle$} is found in the delocalized phase to deviate substantially from the exponential decay. However, data from different values of $\Delta$ all collapse onto a single curve in the localized phase, where the tunneling amplitude seems to be  irrelevant.

In Fig.~\ref{f8}(b), the behavior of the magnetization $\langle \sigma_z \rangle$ is studied under bias values $\varepsilon=1\times 10^{-5}, 1\times10^{-6}, 1\times10^{-7}$, and $1\times10^{-10}$. Other parameters $\Delta=0.1, s=1$, and $\omega_c=1$ are set. With increasing $\varepsilon$, the abrupt jump in $\langle \sigma_z \rangle$ occurring at the transition point is progressively replaced by a smooth behavior. Interestingly, the position where the magnetization falls to $\langle \sigma_z \rangle = -1$ is almost the same, since all the bias values presented here are smaller than the renormalized tunneling strength $\Delta_r \approx 1.9\times 10^{-5}$ at $\alpha_c=0.33$.
For stronger bias or weaker tunneling, however, there is a suppression of the $\alpha$ bracket with $|\langle \sigma_z \rangle|$ smaller than unity. An example is the case of $\Delta=0.01$ in Fig.~\ref{f8}(b) wherein $\varepsilon=1\times10^{-5}$ is larger than $\Delta_{\rm r}(\alpha_{\rm c}) \approx 2.5 \times 10^{-6}$.
Dashed lines provide good fits to the numerical data with the form $y=a/\left[a+\exp(bx)\right]-1$, yielding $b=52.1(5)$ and $45.8(6)$ for $\Delta=0.01$ and $0.1$, respectively.
{\color{red} Inset shows the Kondo energy $T^*$ estimated from the magnetization $\langle \sigma_z \rangle$ in the biased cases \cite{hur08}.
The linear behaviors of the curves imply an exponential scaling $\ln(T^*/\varepsilon) \propto 1/(\alpha_c - \alpha)$ in the delocalized phase for the KT transition.}

We then demonstrate the phase diagram of the two-impurity model with Ohmic dissipation in Fig.~\ref{f9}, in comparison with QMC and NRG results estimated from Res.~\cite{ort10,win14}. In the inset, the transition boundary of NVM shows weak $\Delta$-dependent behavior, and locates at $\alpha_c \approx 0.31+\mathcal {O}(\Delta/\omega_{\rm c})$, much greater than $\alpha_c \approx 0.22$ and $0.18$ in the presence of $\Delta=0.025$ for QMC and NRG, respectively, which could not be ruled out by numerical errors. The underestimation of the transition point in NRG and QMC results may be due to {\color{red}a lack of the convergence toward the continuum, the influence of an imposed bias, and the trapping in the metastable state}. Besides, the phase diagram in the $\alpha$-$\varepsilon$ plane is also investigated (not shown), which exhibits the same qualitative features as that in the $\alpha$-$\Delta$ plane when the bias is small.

The contribution of the spin-spin coupling $K$ to the quantum transition is also presented in Fig.~\ref{f9}. For ferromagnetic case ($K<0$), the phase boundary depends on $K$ very weakly, while for the antiferromagnetic one ($K>0$), $\alpha_c$ increases rapidly with $K$ following the asymptotic line of $\alpha_c=0.25K$, a result consistent with that given by the vanishing renormalized Ising coupling, $K_r=K-4\alpha\omega_c^s/s=0$, as marked by the dashed line in Fig.~\ref{f9} \cite{ort10, zhe15}. The NRG and NVM results for $K \geq 1.5$ concur within the error bars, but the two differ in the ferromagnetic regime where the critical coupling $\alpha_c$ is found to be $0.31(1)$ for NVM and $0.17(1)$ for NRG. It follows that
that metastable states seem to be suppressed at a large $K$, and the phase transition is governed by the interplay between the spin-spin coupling and environment dissipation.

\begin{figure}[tbp]
\centering
\subfigure[]{
\begin{minipage}[b]{0.35\textwidth}
\includegraphics[scale=0.35]{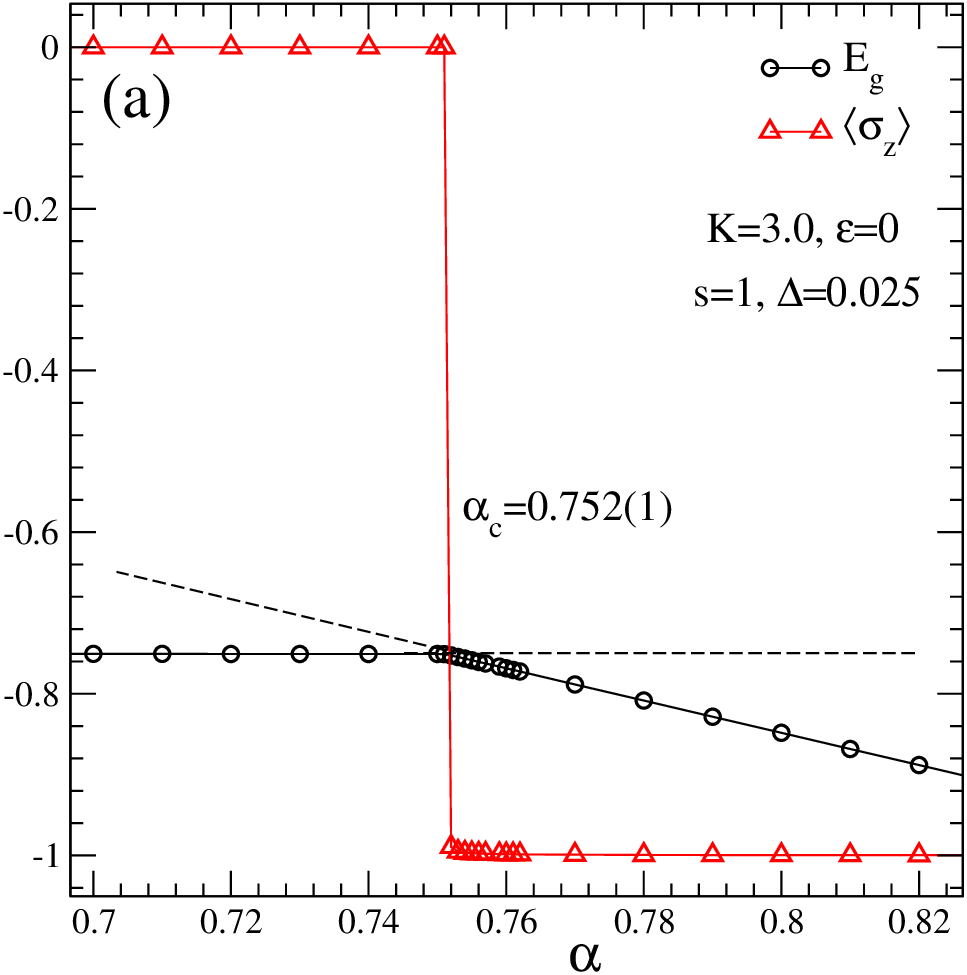}	
\end{minipage}
}
\subfigure[]{
\begin{minipage}[b]{0.35\textwidth}
\vspace{1.5\baselineskip}
\includegraphics[scale=0.35]{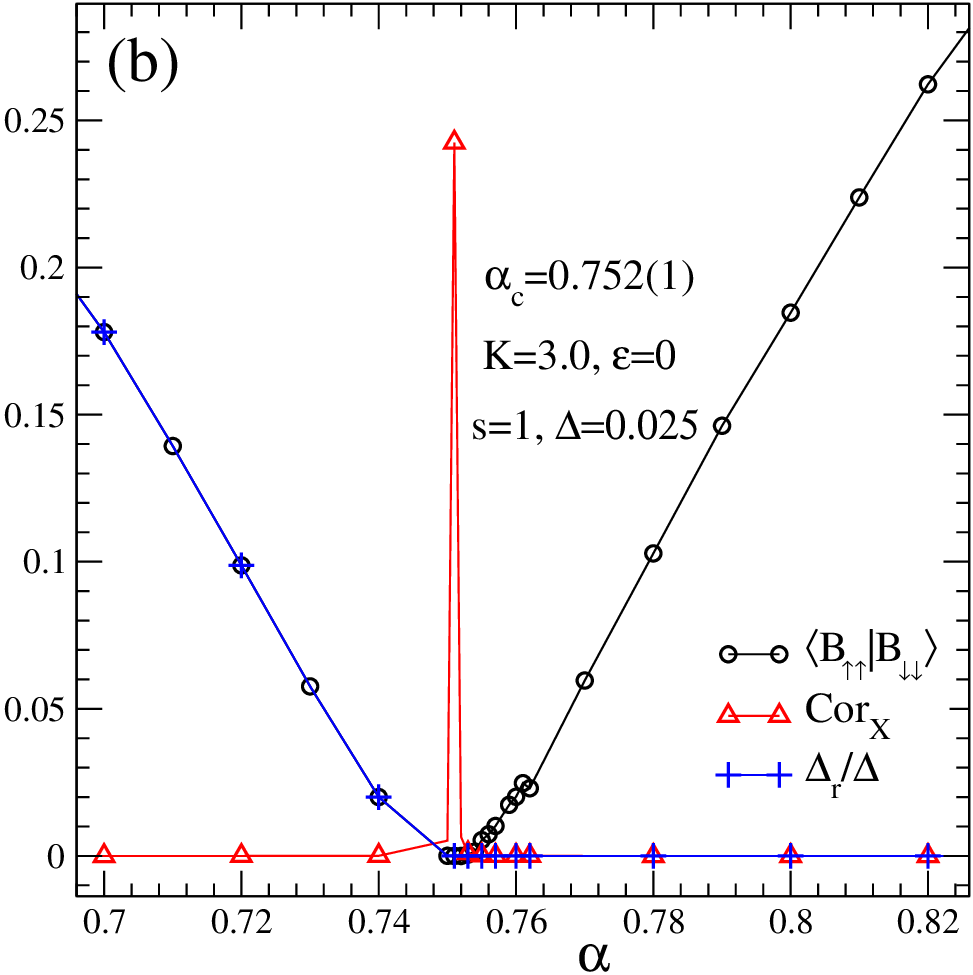}	
\end{minipage}
}
\vspace{-2.5\baselineskip}
\caption{ {\color{red}Ground-state properties including $E_{\rm g},~\langle \sigma_z\rangle,~\Delta_{\rm r}/ \Delta$, $\langle \rm{B}_{\uparrow\uparrow}|\rm{B}_{\downarrow\downarrow}\rangle$, and $\rm Cor_{X}$ are presented as a function of $\alpha$ for the antiferromagnetic case with $K=3.0$. The dashed lines in (a) are guides to the eye for linear fits.}
}
\label{f10}
\end{figure}

Furthermore, we focus on the ground-state properties of the two-impurity model at strong antiferromagnetic coupling, taking $K=3.0$ as an example. The discontinuity in the magnetization $\langle\sigma_z \rangle$ is shown in Fig.~\ref{f10}(a), similar with that in Fig.~\ref{f1}(b). However, two different slopes of $E_{\rm g}$ from linear fittings, i.e., $0.00$ and $2.00$, indicate that the derivative of the free energy with respect to $\alpha$, $\partial E_{\rm g}/ \partial\alpha$, is discontinuous at the transition point, suggesting that the transition is of first order, instead of the Kosterlitz-Thouless type. In Fig.~\ref{f10}(b), {\color{red} both the correlation between two bases $\langle \rm{B}_{\uparrow\uparrow}|\rm{B}_{\downarrow\downarrow}\rangle$ and the correlation function $\rm Cor_X$ are symmetric about the transition point. This points to the same properties of the Ohmic bath in the delocalized and the localized phase, therefore lending further support to the first-order transition. While $\Delta_r/\Delta$ vanishes again in the localized phase. For weak antiferromagnetic coupling, such as $K=0.5$ or $1.0$, the critical couplings are much greater than the $\alpha$ value at which the slope of $E_g$ curve is changed abruptly, suggesting that the transition is impossible to be of first order. Further studies show that it still belongs to Kosterlitz-Thouless universality class.}

{\color{red} Let us turn to the influence of the spin-spin coupling. In the antiferromagnetic regime ($K>0$), the absence of the Kosterlitz-Thouless transition is triggered by the emergence of the antiparallel spin configuration $\langle\sigma_1^z\sigma_2^z\rangle \approx~-1$ in the delocalized phase. The dominant spin-spin coupling term $K \sigma_1^z\sigma_2^z/4$ then leads to an $\alpha$-independent ground-state energy of $E_g \approx -K/4$ and the vanishing of the quantum correlation between bath modes, quite different from the situation at $K=0$. In the localized phase, however, one obtains the parallel state with $\langle\sigma_1^z\sigma_2^z\rangle \approx 1$ instead.

Hence, the energy shift induced by the spin-spin coupling term is $\delta_1=K/4-(-K/4)=K/2$ after the spin-bath coupling strength crosses $\alpha_c$. Substituting the classical displacement coefficients $f_k \approx \lambda_k / \omega_k$, the environmental dissipation term can be simplified as $(\sigma_1^z+\sigma_2^z) \int J(\omega)d\omega/\omega \propto \alpha(\sigma_1^z+\sigma_2^z)$. According to the jump in the magnetization from $\langle\sigma_z\rangle = 0$ to $-1$ at $\alpha_c$ as presented in Fig.~\ref{f10}(a), and the relation $\sigma_1^z=\sigma_z^2=\sigma_z$, the energy shift is calculated as $\delta_2=-2\alpha$, which should be equal to $-\delta_1$ since the ground-state energy $E_g$ and the spin coherence $\langle \sigma \rangle$ are continuous at the transition point. Thus, the transition point $\alpha_c = K/4$ is estimated, consistent with our numerical results in Fig.~\ref{f9} for a large antiferromagnetic coupling strength $K$. Moreover, the ground-state energy is predicted to be $E_g \approx -2\alpha + K/4$ in the localized phase and $-K/4$ in the delocalized phase, in excellent agreement with that in Fig.~\ref{f10}(a). The transition is then inferred to be of first order according to the size of discontinuity in the derivative of the ground-state energy and the few-body nature induced by the independence of bath modes in the two phases. In contrast, the transition is of the Kosterlitz-Thouless type in the ferromagnetic regime ($K<0$), and both the transition point and the critical exponents are almost the same as those in the absence of spin-spin coupling.}

\section{Discussion and conclusions}

As demonstrated by the NVM results in Figs.~\ref{f1}-\ref{f7}, we have accurately determined the transition point {\color{red} $\alpha_c=0.316(8)$} for the two-impurity Ohmic SBM with a tunneling constant of $\Delta=0.025$ in the absence of spin-spin coupling and the bias. Our result is closely supported by the ED result of $\alpha_c\approx0.26$,
but is clearly much greater than the NRG result of $\alpha_c\approx 0.18$ \cite{ort10}. The NRG deviation is likely caused by the minute bias field $10^{-8}\omega_c$ imposed on the NRG calculations. As a consequence, the jump in the magnetization is now replaced by the smooth decay towards $\langle \sigma_z\rangle =-1$ as $\alpha$ is increased to $\alpha_c$ [cf.~Fig.~\ref{f8}(b)]. It follows that in such a situation the transition point should be located by the emergence of $|\langle \sigma_z\rangle| = 1$ instead of a nonzero $|\langle \sigma_z\rangle|$. With this criterion, $\alpha_c \approx 0.3$ can be obtained according to the NRG results in Fig.~\ref{f4} of Ref.~\cite{ort10}, in good agreement with our NVM result here. Similarly, the underestimated result of $\alpha_c=0.125$ reported in Ref.~\cite{zhe15} can also be attributed to a bias of $10^{-5}\omega_c$. Considering the notorious difficulty associated with numerical studies of the Kosterlitz-Thouless transition thanking to the existence of numerous metastalbe states in the vicinity of the ground state, the results of $\alpha_c=0.22$ and $0.16$ from QMC and quantum dynamics simulations in Refs.~\cite{win14,hen16}, respectively, may be unreliable, let alone that of $\alpha_c=0.5$ obtained by the variational treatment based on Silbey-Harris ansatz and simple approximations \cite{mcc10}.

In addition, the quantum criticality of the Ohmic bath has also been investigated. For $\alpha \leq \alpha_c$, perfect power-law behaviors of the  correlation function $\rm Cor_{X}$ and the coordinate variance function of $2\Delta X_{\rm b}-1$ with respect to the frequency $\omega_k$ have been revealed, as shown in Figs.~\ref{f5}(a) and \ref{f6}(a), pointing to a divergent correlation length both in the delocalized phase and at the critical point. Similarity can be drawn with the classical two-dimensional XY model in which the low-temperature phase exhibiting quasi-long-range order is characterized by a power-law behavior of the spin-spin correlations decaying with the distance \cite{kos74}. Analogous to the universal jump of the superfluid density in the XY model \cite{har97}, abrupt changes at the critical coupling take place in the magnetization $\langle \sigma_z\rangle$, correlation function $\rm Cor_{X}$ between the two bath modes at $l=0$ and $k=1$, and departure from the uncertainty relation of $\Delta X_{\rm b}\Delta P_{\rm b} = 1/4$ for the $0$-th bath mode. Combining with the continuous character in derivatives of $\rm E_g$ of any order, it can be concluded that the quantum phase transition of the two-impurity Ohmic SBM belongs to the Kosterlitz-Thouless universality class, similar to that of the two-dimensional XY model.

In summary, the ground-state phase transitions in the two-impurity SBM coupled to a common Ohmic bath have been studied comprehensively by variational calculations using the multi-D$_1$ ansatz. With more than ten thousand variational parameters, the ground-state energy $E_{\rm g}$, spin magnetization $\langle \sigma_z \rangle$, and spin coherence $\langle \sigma_x \rangle$  as well as observables related to the Ohmic bath have been investigated, and {\color{red} the NVM results with the linear discretization show superior accuracy, in comparison with logarithmic-grid results from ED and variational calculations.} A critical coupling strength of $\alpha_c\approx 0.31(1)$ is determined in the weak tunneling limit, which is comparable to the ED result $\alpha_c\approx 0.26$, but very different from results previously obtained by NRG ($0.18$), QMC ($0.22$), and other numerical studies ($0.5, 0.16$, and $0.125$) \cite{ort10, win14, mcc10,hen16, zhe15}. {\color{red} The underestimation of the transition point may be caused by the incorrect criteria of the transition point in the biased case, the lack of the convergence to the continuum, and the existence of the meatstsatble states.} In the ferromagnetic coupling regime $K<0$ and in the absence of spin-spin coupling ($K=0$),  the transitions are found to belong to the  Kosterlitz-Thouless universality class. In the antiferromagnetic regime ($K>0$), however, the transition is inferred to be of first order. Furthermore, we have also examined the influences of the tunneling constant $\Delta$ and the bias field $\varepsilon$, and have established the phase diagram.

\section*{Acknowledgments}
The authors thank Zhe Sun for useful discussions. This work was supported in part by Natural Science Foundation of Zhejiang Province of China under Grant No. LY17A050002.

\appendix

\providecommand{\noopsort}[1]{}\providecommand{\singleletter}[1]{#1}%
\providecommand{\WileyBibTextsc}{}
\let\textsc\WileyBibTextsc
\providecommand{\othercit}{}
\providecommand{\jr}[1]{#1}
\providecommand{\etal}{~et~al.}

\end{document}